\documentclass{JFM-FLM_Au}

\usepackage{bm}
\usepackage{subcaption}
\usepackage[section,above]{placeins}
\usepackage{tikz}
\usepackage{xcolor,booktabs}
\usepackage{textcomp}
\usepackage{xcolor}
\usepackage{natbib}

\makeatletter
\AtBeginDocument{%
  \def\oddabsfooterflag{\hbox to \hsize{\hfil{\cppagefont\thepage}}}%
  \def\evenabsfooterflag{\hbox to \hsize{{\cppagefont\thepage}\hfil}}%
  \def\pagelimitfooter{\hbox to \textwidth{{\cppagefont\thepage}\hfill}}%
  \def\ps@titlepage{\leftskip\z@\let\@mkboth\@gobbletwo
    \def\@oddhead{\hbox to \textwidth{\@j@urnal\hfill}}\let\@evenhead\@oddhead
    \let\@oddfoot\oddabsfooterflag\let\@evenfoot\evenabsfooterflag
    \def\sectionmark##1{}}%
  \def\ps@headings{\let\@mkboth\markboth
    \def\@oddhead{\hfill{\itshape\@righttitle}\hfill}%
    \def\@evenhead{\hfill\itshape\@lefttitle\hfill}%
    \def\@oddfoot{\hbox to \textwidth{\hfil{\cppagefont\thepage}}}%
    \def\@evenfoot{\hbox to \textwidth{{\cppagefont\thepage}\hfil}}%
    \def\sectionmark##1{\markboth{##1}{}}%
    \def\subsectionmark##1{\markright{##1}}}%
  \pagestyle{headings}}
\makeatother

\lefttitle{Pandit et al.}
\righttitle{Journal of Fluid Mechanics}

\title{Flow localization and rarefied gas flow in hierarchical porous aerogels}

\author{Prakul Pandit\aff{1}, Rucha Kunte\aff{1}, Marina Schwan\aff{1}, Cedric J. Gommes\aff{2}, Julia Kowalski\aff{3} \and Ameya Rege\aff{1,4}}

\affiliation{\aff{1}German Aerospace Center (DLR), Institute for Frontier Materials on Earth and in Space, Linder Höhe, 51147 Cologne, Germany
\aff{2}University of Liège, Department of Chemical Engineering, B6A, Allée du Six Août 3, 4000 Liège, Belgium
\aff{3}RWTH Aachen University, Faculty of Mechanical Engineering, Eilfschornsteinstraße 18, 52062 Aachen, Germany
\aff{4}University of Twente, Department of Mechanics of Solids, Surfaces \& Systems, P.O. Box 217, 7500 AE Enschede, The Netherlands}

\corresau{Ameya Rege, \email{ameya.rege@utwente.nl}}

\begin{document}

\maketitle
\begin{abstract}

Many aerogels exhibit a hierarchically porous morphology comprising of interconnected meso- and macropores. Gas transport in such microstructures occurs predominantly in the slip and transition regimes, where rarefaction effects are coupled to the multiscale pore geometry. This resulting structural heterogeneity challenges conventional analytical descriptions of permeability based on simplified geometric representations and a single characteristic length. In this paper, we investigate rarefied gas transport in resorcinol-formaldehyde aerogels by reconstructing their pore morphology using Gaussian random fields, with the reconstructed microstructures calibrated against bulk density, nitrogen sorption and electron microscopy. The resulting structures are converted into pore networks for gas transport simulations incorporating the Beskok-Karniadakis rarefaction correction at the individual throat level. We show that mesopore-scale reconstructions underestimate experimentally measured permeabilities by more than an order of magnitude, whereas the proposed two-field Gaussian random field framework, explicitly accounting for hierarchical porosity, substantially improves predictive accuracy. Furthermore, gas transport is found to be concentrated within a sparse backbone of macropore throats, resulting in a flux-weighted Knudsen number more than an order of magnitude lower than its number-weighted counterpart. Bounds derived from the minimum-dissipation principle reveal that rarefaction-induced permeability enhancement is governed by a dissipation-weighted characteristic throat radius that exceeds the sorption-derived characteristic pore radius by more than an order of magnitude. Consequently, sorption-based estimates systematically overpredict rarefaction-induced permeability enhancement by attributing transport to surface-area-dominant mesopores rather than transport-controlling macropore throats. These findings establish the importance of distinguishing between structurally representative and transport-relevant length scales, providing a physical basis for permeability predictions in hierarchically porous media.

\end{abstract}


\section{\label{sec:intro}Introduction}

Organic aerogels are meso-porous materials synthesized by sol-gel polymerization of multifunctional organic monomers followed by supercritical drying \citep{pekala1993structure, garcia2025review}. In resorcinol-formaldehyde (RF) aerogels, an interconnected network of primary particles forms a hierarchical pore space spanning the micro-, meso-, and macropore regimes \citep{Pekala1989}. A fundamental property governing the performance of these RF aerogels in transport-dependent applications is their permeability. Whether determining the infiltration into carbon aerogel electrodes \citep{nojabaee2021ultramicroporous}, the convective component of heat transfer in an insulating panel, or the rate of solvent exchange during supercritical drying, the ease with which a fluid can traverse the aerogel network is a decisive performance metric. However, accurately characterizing the permeability of these high-porosity amorphous systems remains inherently challenging. Experimental measurements are complicated by the material's mechanical fragility. Even small mechanical loads, in particular through shear forces, can lead to fracture or plastic deformation that might invalidate measurements. Furthermore, no standardized commercial apparatus exists for measuring gas permeability through mesoporous materials, and custom setups require careful, pressure-tight mounting that has a significant impact on results~\citep{Beyerlein}. \citet{Anez2014} measured both gas and liquid permeability in nanocomposite silica aerogels, reporting that gas permeability exceeds liquid permeability by more than an order of magnitude, a discrepancy attributed to the slip (Klinkenberg) and transition (Knudsen) flow regimes that arise when the molecular mean free path is comparable to the pore diameter. \citet{Beyerlein} measured gas permeabilities of $10^{-15}$-$10^{-13}$~m$^{2}$ in polyurethane aerogels with a dynamic pressure method. These measurements are usually interpreted with a Knudsen number based on a single pore size, but in hierarchical aerogels this size is ambiguous. Mesopores resolved by nitrogen sorption carry most of the surface area, whereas macropores outside the sorption range hold most of the pore volume~\citep{Reichenauer2001}, and it is not known which of the two sets the rarefaction of the flow. Lattice Boltzmann methods (LBM) have been widely used for permeability prediction in porous materials~\citep{Pan2006,Chen1998} and can, in principle, resolve arbitrary pore geometries at the voxel level. In hierarchical aerogels, however, a representative volume must contain macropores of several hundred nanometres while resolving mesopores of a few nanometres, and rarefied-flow simulations at this resolution remain too costly for systematic parametric studies. Kinetic simulations in simplified and digitized porous media have shown that slip-flow corrections can misrepresent the apparent permeability in the transition regime~\citep{Wu2017} and that the largest flow paths dominate the rarefied permeability~\citep{Ho:2019}. An alternative approach is provided by pore network models (PNM), which reduce the pore space to a topologically equivalent graph of pores and throats~\citep{gostick2017versatile}. PNMs have been successfully applied to predict gas permeability in shale and tight formations, where \citet{Mehmani2013} added slip flow and Knudsen diffusion to the viscous conductance of a multiscale network of pores and obtained the pressure dependence of the apparent permeability of shale. This framework has since been adopted for various nanoporous geomaterials, and network models have also been used to describe gas-phase and solid conduction in open-porous nanomaterials~\citep{aney2025network}, yet rarefied gas flow has not been treated in the high-porosity ($\phi_0 > 0.7$), bicontinuous morphologies characteristic of organic aerogels. For shales, apparent-permeability models that combine the pore structure with slip and Knudsen transport~\citep{javadpour2009,Darabi2012} have become the established route from pore structure to gas permeability.

\begin{figure*}
\centering
\includegraphics[width=\textwidth]{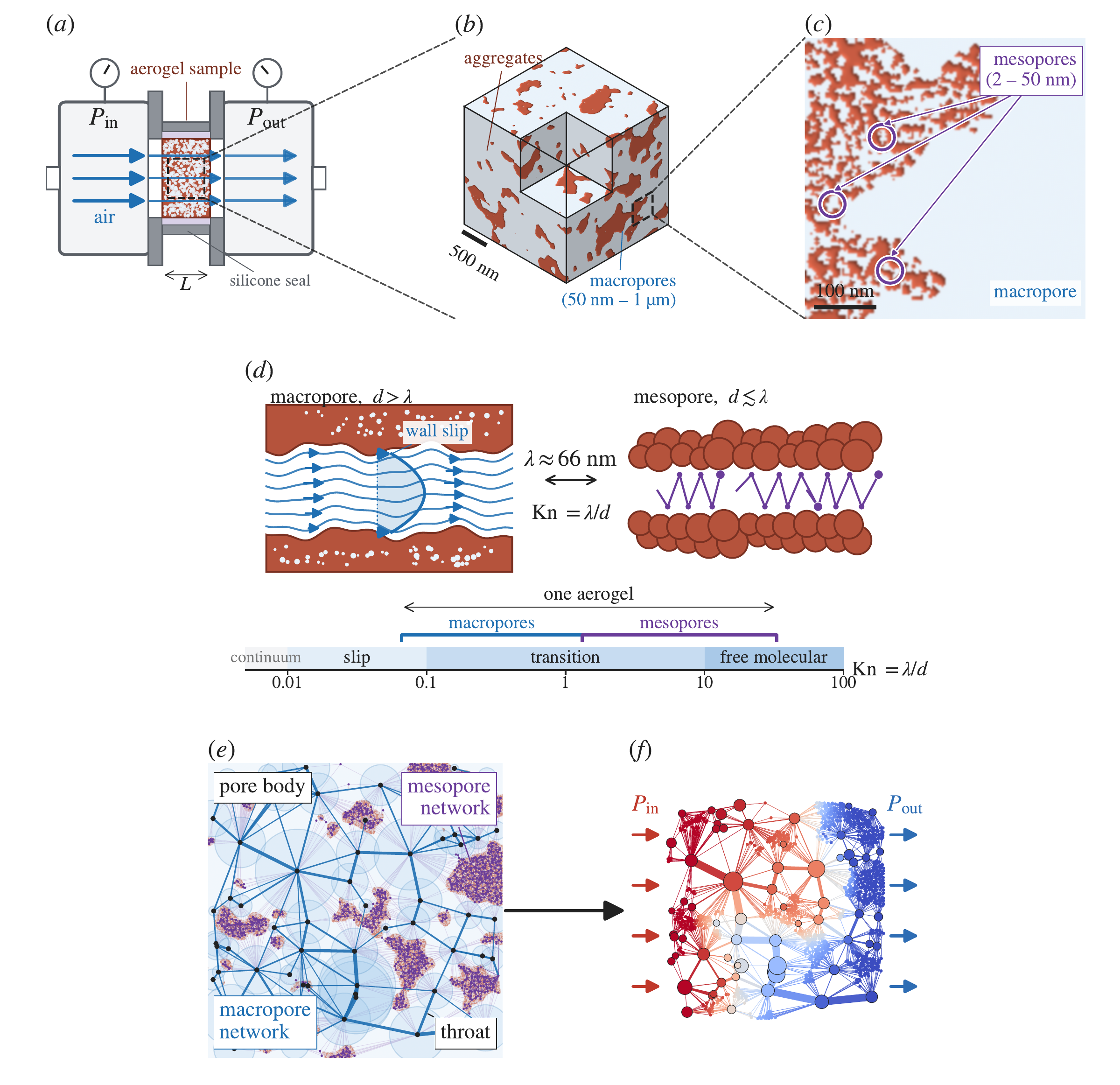}
\caption{\label{fig:overview}
Conceptual overview of gas transport in a hierarchically porous aerogel. ($a$) Permeability test setup with the sample sealed between two plates. ($b$) Cut-away of the reconstruction of sample 7A with the proposed model and ($c$) the mesoporous architecture inside one aggregate. ($d$) Viscous flow with wall slip in a macropore and molecule-wall transport in a mesopore, with the two pore classes of one sample placed on the Knudsen axis. ($e$) Pore network extracted from the macropore space and ($f$) the steady gas flow solved on it, with node colour showing pressure and line width the throat flow.}
\end{figure*}

In low-porosity materials, a pore network can often be formulated in terms of individual pores embedded in a continuous solid matrix. However, in high-porosity aerogels (\(\phi_0 > 0.7\)), the void forms a continuous and highly interconnected space, while the solid skeleton occupies only a small fraction of the total volume. The definition of individual pore bodies is therefore less straightforward, and the distinction between pores and the constrictions connecting them (pore throats) becomes dependent on the microstructure morphology. Previous studies of liquid permeability in wet gels have addressed this geometry by modelling the solid framework as networks of cylinders \citep{scherer1994}, an approach also applied to RF gels \citep{scherer1996}. Predicting gas transport in aerogels additionally requires accounting for rarefaction effects across the transition from continuum to free-molecular flow. A suitable modelling framework must therefore represent the interconnected pore geometry, particularly its constrictions, as well as the relevant transport physics. To this end, the present work investigates gas transport through hierarchically porous RF aerogels using a computational framework complemented by experimental permeability measurements. Figure~\ref{fig:overview} provides a conceptual overview of the study. Figure 1($a$) presents a schematic of the experimental setup, in which an RF aerogel specimen is positioned within a gas chamber and sealed at the top and bottom to prevent bypass flow. Figures 1($b$) and ($c$) illustrate the hierarchical pore architecture of RF aerogels, while Figs. 1($c$) and ($d$) depict the distinct gas transport mechanisms operating at the different characteristic length scales. Distinguishing between these length scales is fundamental to understanding gas transport in such materials. Aggregates of primary particles form an interconnected network of macropores with characteristic dimensions of several hundred nanometres, whereas mesopores with dimensions of a few nanometres reside within the aggregates. At atmospheric pressure and room temperature, the mean free path of air molecules is approximately $66~\mathrm{nm}$, lying between the characteristic dimensions of these two pore populations. Consequently, gas transport through the macropores is governed by viscous flow with non-negligible rarefaction effects, whereas molecule-wall collisions become increasingly dominant within the mesopores, where transport may approach the Knudsen regime. The hierarchical pore architecture thus gives rise to the coexistence of distinct gas transport regimes within a single material. The macroscopically measured permeability consequently emerges from the interplay between the local transport mechanisms, the spatial organisation of the pore populations, and their connectivity across length scales. Identifying the dominant flow pathways and elucidating the individual and collective contributions of rarefaction effects at different length scales constitute central objectives of the present study.

Given the hierarchical architecture of the investigated aerogels and the inherent challenges associated with experimentally reconstructing their three-dimensional microstructures at sub-100 nm length scales~\citep{tannert2017three}, a stochastic computational approach is adopted to generate representative microstructures. Specifically, we employ a methodology based on level-cut Gaussian random fields (GRFs), considering two distinct microstructural representations: a conventional single-field model, following earlier studies~\citep{Gommes:2018, gommes2008structure}, and a newly proposed two-field model. The latter explicitly incorporates the hierarchical nature of the pore architecture through the intersection of two statistically independent level-cut GRFs associated with the characteristic length scales of the mesopore and macropore populations. This framework enables a systematic investigation of the influence of hierarchical pore morphology on gas transport and the resulting effective permeability. The parameters of the reconstructions are calibrated against bulk density, nitrogen sorption and electron microscopy data. The connectivity of the pore space and its partition into pore bodies and connecting throats are determined directly from the generated geometry. The reconstructed structures are then converted into pore networks using SNOW2~\citep{gostick2019porespy}, an implementation of the SNOW algorithm~\citep{gostick2017versatile}. Gas transport is modelled using Hagen--Poiseuille conductance, corrected at each throat using the Beskok--Karniadakis formulation. Figures ~\ref{fig:overview}  ($e$) and ($f$) show the extracted network and the resulting pressure field, respectively. No reconstruction parameters are fitted to permeability data; therefore, comparison with measurements on two RF aerogels provides an independent test of the structural models. The flux-resolved network solutions show that flow concentrates in a sparse backbone of macropore throats. We further use the minimum-dissipation principle for networks~\citep{DoyleSnell:1984} to derive bounds on the resulting rarefaction enhancement. These bounds identify a single dissipation-weighted throat radius governing the enhancement, linking the critical-path picture of transport~\citep{ambegaokar1971,katz1986} to the transition regime and providing a basis for recovering the intrinsic permeability from a gas-permeability measurement at a single pressure.

The remainder of the paper is organized as follows. Section~\ref{sec:exp} describes the synthesis and experimental characterization of the RF aerogels, including the measured permeabilities and pore-volume balance. Section~\ref{sec:comp} presents the GRF-based reconstruction and characterizes the resulting microstructures. Section~\ref{sec:results} describes the transport models, compares their predictions with the measurements, and analyses flow localization and the bounds on the rarefaction enhancement. Section~\ref{sec:conclusions} summarizes the principal findings.

\section{\label{sec:exp}Experimental analysis}

\FloatBarrier
\subsection{\label{sec:exp:synthesis}Synthesis of resorcinol-formaldehyde aerogels}
At room temperature, resorcinol (R) (98\%, Aldrich) was dissolved in deionized water (W) and stirred at 150~rpm using a cross-shaped magnetic stirring bar. Subsequently, an aqueous solution of formaldehyde (F) (23.5\% w/w, stabilized with 1\% methanol, Carl Roth) and sodium carbonate (C) (Aldrich) was added to the stirred resorcinol solution. The molar ratios of the chemicals, identical for all samples, are listed in Table~\ref{tab:synthesis}. The pH of the final solution was adjusted to 6.5 with 2~molar nitric acid (PanReac AppliChem).
Stirring at room temperature was continued, after which the homogeneous and transparent solution was transferred into a sealable polypropylene container and stored in an oven at 60~\textdegree C (Memmert GmbH, Germany) for 7~days.

After the specified gelation time, the wet gels were cooled to room temperature and transferred into an acetone bath (pure, technical grade, TH.~Geyer) to remove residual chemicals and to exchange water with acetone, which is soluble in supercritical carbon dioxide. The acetone exchange was performed with a pump operated at 150~rpm. For sample 7A, zeolites (Thermo Fisher Scientific) were additionally used to remove residual water (Table~\ref{tab:synthesis}). Supercritical drying was carried out using CO$_2$ (purity $\geq$99.995\%, Praxair) in a 60~L autoclave (Eurotechnica, Germany) at 60~\textdegree C and 110~bar for approximately 21~h. The depressurization rate was controlled at 0.2~bar~min$^{-1}$.

\begin{table}
\begin{center}
\begin{tabular}{lccccc}
\hline
Sample & R/W & R/C & R/F & Washing & Shrinkage (\%) \\
\hline
6B & 0.014 & 200 & 0.5 & pump & 8.5 \\
7A & 0.014 & 200 & 0.5 & pump + zeolites & 8.0 \\
\hline
\end{tabular}
\caption{Synthesis parameters of the RF aerogel samples:
molar ratios of resorcinol (R), water (W), catalyst (C),
and formaldehyde (F), washing procedure during solvent
exchange, and linear shrinkage after supercritical drying.}
\label{tab:synthesis}
\end{center}
\end{table}

\subsection{\label{sec:exp:characterization}Experimental characterization and pore-size analysis}

\paragraph{Density measurements}
The bulk densities of the samples were measured using a GeoPyc instrument, while the skeletal densities were determined using an AccuPyc instrument. Both measurements are based on displacement techniques (Micromeritics, Germany).

\paragraph{Physisorption}
Prior to physisorption measurements, the samples were degassed for 12~h at 100~\textdegree C under a vacuum of 0.1-0.5~mbar using a SmartVacPrep system (Micromeritics, Germany). Nitrogen adsorption-desorption isotherms were recorded at 77~K using a TriStar analyzer (Micromeritics, Germany) to determine the specific surface area, the mesopore size distribution, and the mesopore volume. The measured sample properties are summarized in Table~\ref{tab:sample_props}.

\begin{table}
\begin{center}
\begin{tabular}{lcccc}
\hline
Sample 
& $\rho_{\mathrm{env}}$ (g\,cm$^{-3}$) 
& $\rho_{\mathrm{skel}}$ (g\,cm$^{-3}$) 
& $S_{\mathrm{BET}}$ (m$^2$\,g$^{-1}$) 
& $V_{\mathrm{BJH}}$ (cm$^3$\,g$^{-1}$) \\
\hline
6B & 0.142 & 1.488 & 318 & 0.80 \\
7A & 0.141 & 1.528 & 345 & 0.61 \\
\hline
\end{tabular}
\caption{Measured properties of the RF aerogel samples: 
envelope and skeletal densities, BET specific surface area, 
and cumulative BJH pore volume integrated over the sorption 
window ($1.9$-$164$~nm) of the measured 
$\mathrm{d}V/\mathrm{d}\log d$ distributions.}
\label{tab:sample_props}
\end{center}
\end{table}

\paragraph{Pore-volume balance}
The densities and sorption data of Table~\ref{tab:sample_props} can be combined into a pore-volume balance that anticipates the central structural issue of this work. The total specific pore volume follows directly from the measured envelope and skeletal densities as
\begin{equation}
    V_{\mathrm{pore}} = \frac{1}{\rho_{\mathrm{env}}} - \frac{1}{\rho_{\mathrm{skel}}},
    \label{eq:vtot}
\end{equation}
which yields $V_{\mathrm{pore}} = 6.39$~cm$^3$g$^{-1}$ for sample 6B and $6.43$~cm$^3$g$^{-1}$ for sample 7A (Table~\ref{tab:sample_props}). In comparison, the cumulative pore volume obtained from nitrogen physisorption over the full BJH analysis range of $1.9$ to $164$~nm is $V_{\mathrm{BJH}} = 0.80$~cm$^3$g$^{-1}$ for sample 6B and $0.61$~cm$^3$g$^{-1}$ for sample 7A. The fraction of the total pore volume resolved by the BJH analysis is therefore
\begin{equation}
    f_{\mathrm{BJH}} = \frac{V_{\mathrm{BJH}}}{V_{\mathrm{pore}}},
    \label{eq:fbjh}
\end{equation}
giving $f_{\mathrm{BJH}} = 0.125$ and $0.095$ for samples 6B and 7A, respectively. Consequently, approximately $87$-$90\%$ of the total pore volume lies outside the pore-size range resolved by the BJH analysis, corresponding to unresolved pore volumes of $5.59$ and $5.82$~cm$^3$g$^{-1}$. These values are consistent with previous observations that nitrogen physisorption captures only a limited fraction of the pore volume in highly porous aerogels, particularly when a substantial macroporous network is present~\citep{Reichenauer2001,bjh_bioaerogels2022}. For comparison with the simulations, the BET surface area is also expressed per unit envelope volume, $S_{\mathrm{bulk}} = \rho_{\mathrm{env}} S_{\mathrm{BET}} = 4.5\cdot10^{7}$~m$^2$m$^{-3}$ for 6B and $4.9\cdot10^{7}$~m$^2$m$^{-3}$ for 7A.

\begin{figure*}
\centering
\includegraphics[width=0.95\textwidth]{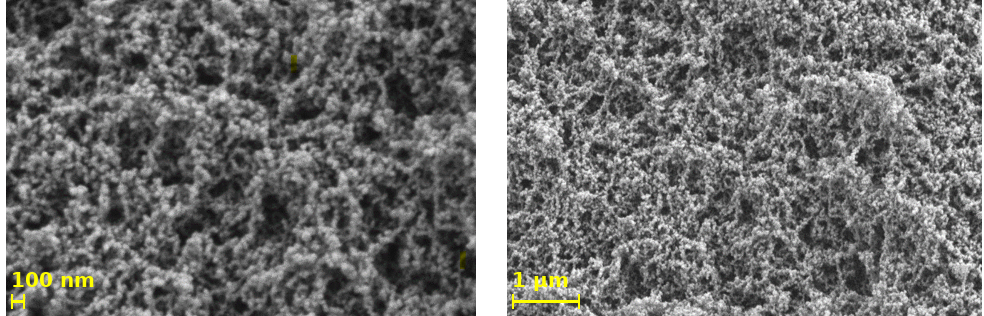}
\caption{\label{fig:sem}
Scanning electron micrographs of RF aerogel sample 6B. The left panel shows the colloidal primary-particle strands and the macropores between them at higher magnification, and the right panel an overview of the aggregate network with inter-aggregate voids.}
\end{figure*}

Where the unresolved 87 to 90\% of the pore volume resides is not resolved by the pore-size analysis, but the BET surface area constrains it. A pore population of volume $V$ whose local width nowhere exceeds $d$ has a surface area of at least about $2V/d$, the value attained by slit-shaped pores, whereas cylinders of diameter $d$ give $4V/d$. Placing the unresolved volume below the lower limit of the BJH analysis ($d < 1.9$\,nm) would therefore require at least $5.9 \cdot 10^{3}\,\mathrm{m^{2}\,g^{-1}}$ (6B) and $6.1 \cdot 10^{3}\,\mathrm{m^{2}\,g^{-1}}$ (7A), more than ten times the measured BET areas. Conversely, since the surface of the unresolved pores cannot exceed $S_{\mathrm{BET}}$, their hydraulic diameter satisfies
\begin{equation}
d_{h,\mathrm{unres}} = \frac{4V_{\mathrm{unres}}}{S_{\mathrm{unres}}} \geq \frac{4V_{\mathrm{unres}}}{S_{\mathrm{BET}}} = 70\,\mathrm{nm}\ \text{(6B)} \quad \text{and} \quad 67\,\mathrm{nm}\ \text{(7A)},
\label{eq:dhbound}
\end{equation}
where $d_h$ is the cylinder-equivalent diameter. The corresponding slit-based bound on the local pore width, $2V_{\mathrm{unres}}/S_{\mathrm{BET}}$, is half this value. Both bounds rest only on the measured densities and BET surface area and are independent of the BJH distribution. They place the unresolved volume at the upper end of the mesopore range and in the macroporous network, consistent with the inter-aggregate voids of one to a few hundred nanometres seen in the electron micrographs (Fig.~\ref{fig:sem}). These SEM images were recorded with a Zeiss Ultra 55 microscope. Prior to imaging, the aerogels were sputter-coated with platinum for 90~s at 16~mA. Figure~\ref{fig:sem} shows sample 6B at two magnifications. The close-up (left panel) resolves the interconnected primary-particle strands characteristic of RF aerogels, whose interstices form the mesopore population probed by nitrogen sorption, while the overview (right panel) reveals a homogeneous network of particle aggregates interspersed with voids of one to a few hundred nanometres.

The cylindrical samples, with diameters of $4.4$~cm (6B) and $4.5$~cm (7A) and height $15$~mm were sealed with a silicone gel to ensure that no gas bypassed the porous body. The test gas utilized was air. Before each run the downstream chamber was pumped down to a pressure $\Delta p$ below the nominal supply pressure of $1000$~mbar, that is to $940$-$980$~mbar, and its pressure rise was then recorded while air flowed through the sample from the upstream side at ambient pressure. A detailed description of the experimental setup and methodology can be found in \citet{ganesan2019gas}, and the underlying theory of gas permeation measurements on aerogels in \citet{ratke2021aerogels}. Following \citet{Beyerlein}, the permeability is extracted from the pressure relaxation of the downstream chamber. For isothermal Darcy flow of an ideal gas, inserting the density $\rho = pM/(RT)$ into Darcy's law and integrating across the sample height $h$ gives the mass flow rate through the sample as
\begin{equation}
\dot{m} = \frac{k A M}{2\,\mu\,h\,R T}\left(p_f^2 - p_c^2\right),
\label{eq:mdot}
\end{equation}
where $p_f$ is the constant upstream (supply) pressure and $p_c$ the downstream chamber pressure. The quadratic pressure dependence arises because the gas expands along the flow path. Equating $\dot{m}$ to the rate of gas accumulation in the closed downstream volume $V_c$, $\dot{m} = (V_c M/RT)\,\mathrm{d}p_c/\mathrm{d}t$, yields
\begin{equation}
\frac{\mathrm{d}p_c}{\mathrm{d}t} = \frac{k A}{2\,\mu\,h\,V_c}\left(p_f^2 - p_c^2\right),
\label{eq:pcode}
\end{equation}
whose solution for an initial downstream pressure $p_0$ is
\begin{equation}
p_c(t) = p_f\tanh(t/\tau_c + \mathrm{arctanh}(p_0/p_f)) ,
\label{eq:tanhfit}
\end{equation}
where $p_f$ is the plateau toward which the chamber relaxes and $\tau_c$ the relaxation time. The recorded experimental points were fitted with Eq.~\eqref{eq:tanhfit}, and solving the expression for the permeability~\citep{Beyerlein} gives
\begin{equation}
k = \frac{2\,\mu\,h\,V_c}{A\,p_f\,\tau_c} ,
\label{eq:kfromtau}
\end{equation}
where $A$ is the cross-sectional area of the sample, $h$ its thickness, $\mu = 1.81\cdot10^{-5}$~Pa\,s the dynamic viscosity of air at $20$~\textdegree C, and $V_c = 0.014$~m$^3$ the calibrated dead volume of the apparatus. 
\par

\begin{figure*}
\centering
\includegraphics[width=0.92\textwidth]{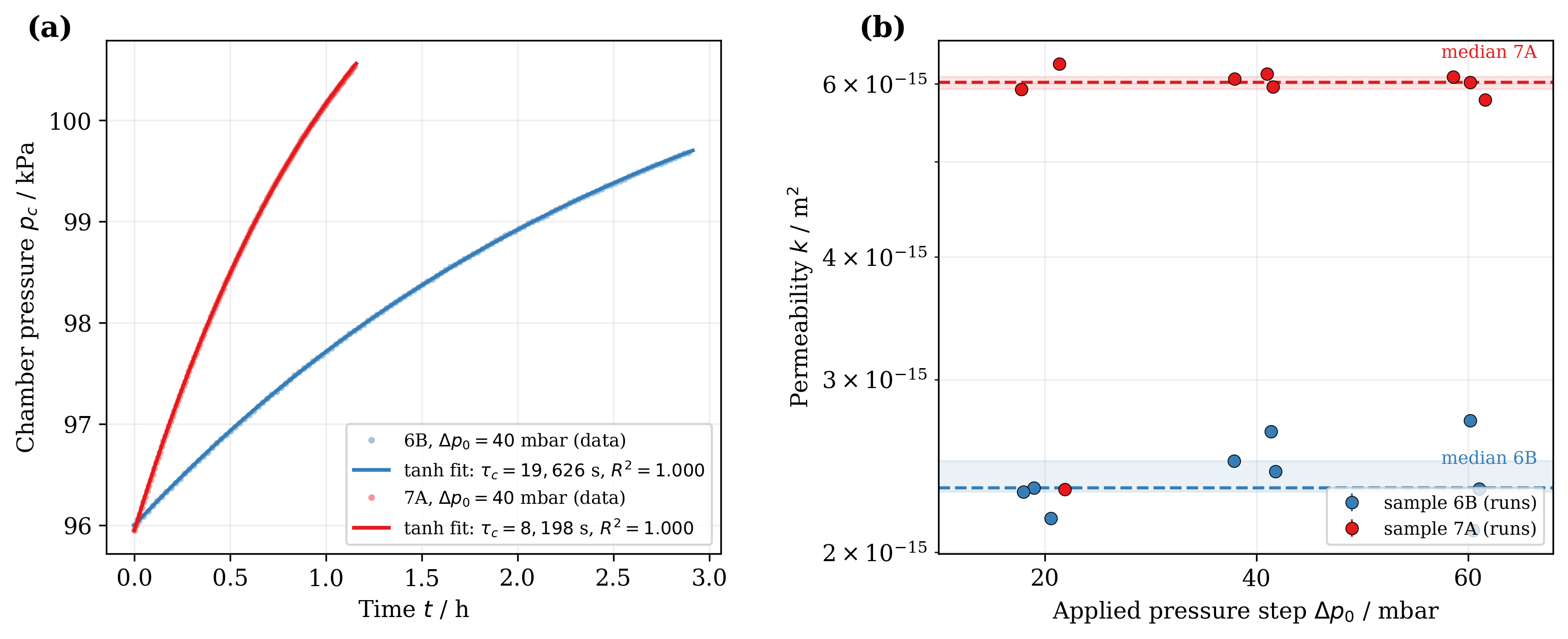}
\caption{\label{fig:exp_permeability}
 Experimental gas permeability of samples 6B and 7A. (a)~Representative pressure-rise measurements with their tanh fits. (b)~Per-run permeabilities versus applied pressure step, with dashed lines and bands marking the per-sample median and interquartile range.}
\end{figure*}

Each sample was measured in nine runs (pressure differences $\Delta p = 20$, $40$, $60$~mbar, three repeats each). The recorded pressure traces are fitted with $p_f$ and $\tau_c$ as free parameters. Runs exhibiting anomalously fast relaxation ($\tau_c < 10^{3}$~s, two orders of magnitude faster than intact runs of the same sample) indicate bypass flow through cracks formed under load and were excluded from the analysis. All retained fits have $R^2 > 0.90$, and the permeability is independent of the applied pressure step to within $8\%$ (6B) and $1\%$ (7A). The resulting permeabilities are $k_{\mathrm{exp}} = 2.26\cdot10^{-15}$~m$^2$ (IQR $2.15$-$2.33\cdot10^{-15}$) for 6B and $5.62\cdot10^{-15}$~m$^2$ (IQR $5.54$-$5.65\cdot10^{-15}$) for 7A. For comparison, the density correlation that \citet{Kong1993} measured with nitrogen near atmospheric pressure on carbon aerogels of the same $\mathrm{R/C} = 200$ recipe, $K = 8.86\cdot10^{-13}\,\rho_{\mathrm{env}}^{-1.48}$~cm$^2$ with $\rho_{\mathrm{env}}$ in g\,cm$^{-3}$, gives $1.6\cdot10^{-15}$~m$^2$ at both densities, below the measured values by factors of $1.4$ (6B) and $3.5$ (7A). Figure~\ref{fig:exp_permeability} shows the highest-$R^2$ pressure-rise trace of each sample together with all per-run values. With the plateau pressure fitted, the nine runs of each sample agree to within a factor of $1.3$ (6B) and, excluding a single outlier, $1.1$ (7A).

\section{\label{sec:comp}GRF-based modelling of aerogel microstructures}

\FloatBarrier
\subsection{\label{sec:comp:grf}Single-field GRF and the proposed model}

Statistically representative digital microstructures of disordered materials can be obtained by level-cutting GRFs~\citep{Berk:1987,Teubner:1991,roberts1996composite,Gommes:2018}, an approach previously applied to gels and aerogels~\citep{roberts1997aerogel,quintanilla2003aerogel}. A statistically homogeneous and isotropic Gaussian field $Y(\mathbf{x})$ of zero mean and unit variance is thresholded at a level $a$, the solid being the region $Y > a$, so that the porosity $\phi_0 = \Pr[Y < a]$ fixes the threshold (Appendix~\ref{app:grf}). The morphology is controlled by the two-point correlation function of the field, for which we use the model introduced by \citet{gommes2008structure} for the small-angle scattering of RF gels,
\begin{equation} \label{eq:gY}
g_Y(r) = \frac{1}{\cosh\left( \frac{r}{\nu \xi} \right)}
\cdot
\frac{\sin\left( \frac{2\pi r}{\xi} \right)}{2\pi r / \xi}.
\end{equation}
with a characteristic length $\xi$ and a dimensionless order parameter $\nu$, wherein large values of $\nu$ yield quasi-periodic structures and small values more disordered ones. Because $g_Y$ is quadratic at the origin, $g_Y(r) \simeq 1 - (r/l_Y)^2$, with the correlation length
\begin{equation}
\label{eq:ly}
\frac{1}{l_Y} = \frac{1}{\nu \xi}
\sqrt{ \frac{1}{2} + \frac{2\pi^2 \nu^2}{3} },
\end{equation}
the level-cut structure has the specific surface area~\citep{Teubner:1991}
\begin{equation}
    S = \frac{2\sqrt{2}}{\pi}\, \frac{e^{-a^2/2}}{l_{Y}},
    \label{eq:S_single}
\end{equation}
which depends on $\nu$ only through $l_Y$. Structures with the same porosity and correlation length therefore have the same surface area for any value of $\nu$. The remaining effect of $\nu$ on pore size is quantified in Appendix~\ref{sec:comp:nueffect}. The limit $\nu \to 0$ is taken at fixed $l_Y$, so all $\nu = 0$ structures are specified by their correlation length (Appendix~\ref{app:grf}).

\begin{figure}
\centering
\includegraphics[width=0.6\textwidth]{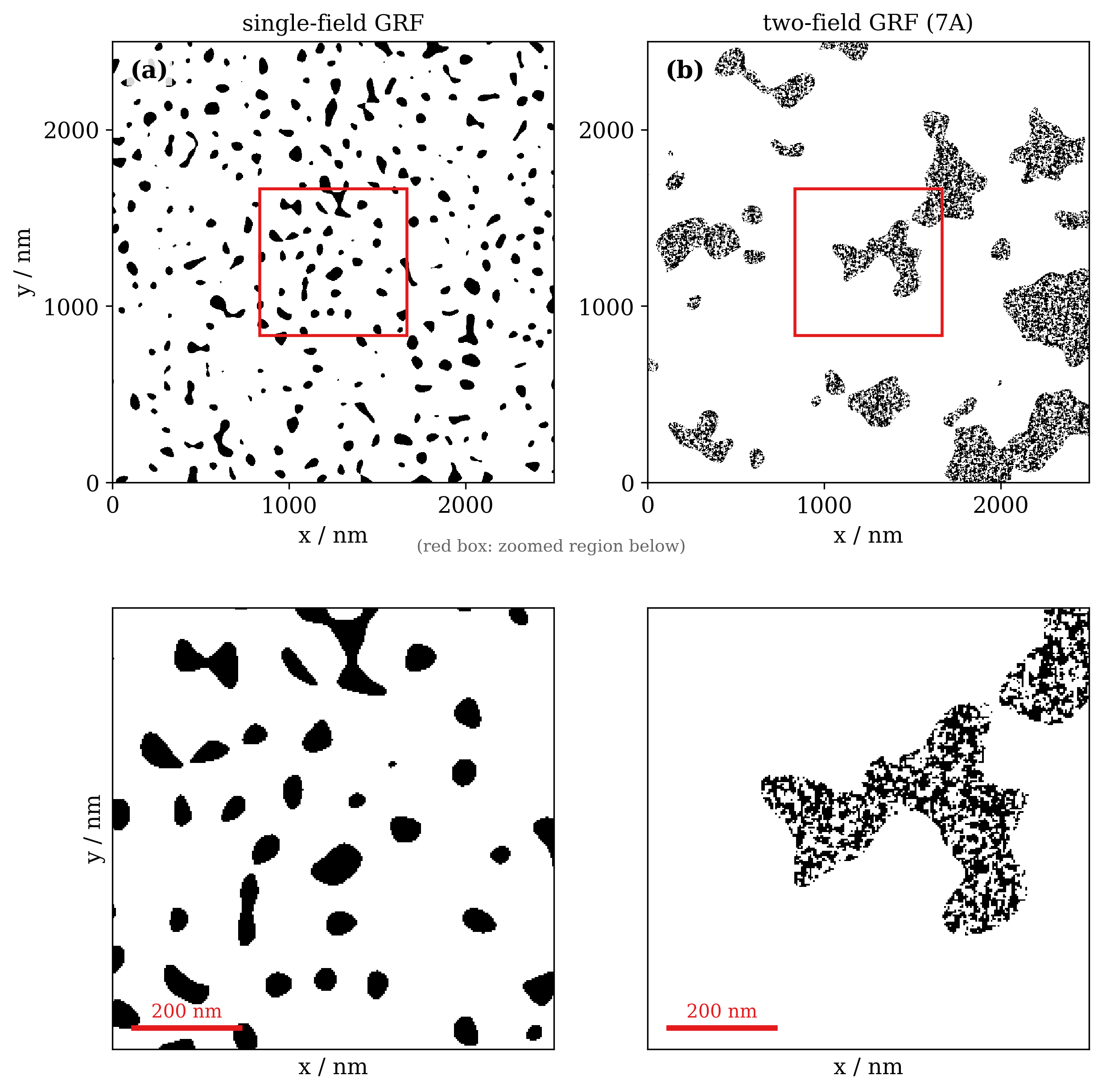}
\caption{\label{fig:grf_xsections}
Cross-sections of (a)~the single-field and (b)~the proposed model at $\phi_0$ of approximately $0.90$ (solid black, void white).}
\end{figure}

A single field has only one structural length, whereas RF aerogels are typically hierarchical. The mesopores resolved by nitrogen sorption coexist with macropores, which hold most of the void volume and, because conductance scales as $d^4$, must contribute most to the flow. If $l_Y$ is calibrated to the measured surface area, all structure falls in the mesopore range and the permeability thus collapses. However, if it is calibrated to the macropores instead, the surface area, which scales as $1/l_Y$, falls more than an order of magnitude below the measured value. We therefore follow the classical approach of intersecting independent structures~\citep{jeulin2012multiscale,Gommes:2018,Gommes:2022} and combine two statistically independent fields at $\nu = 0$. A coarse field $Y_c$ with correlation length $l_{Y,c}$ represents the macropores, and a fine field $Y_f$ with $l_{Y,f} \ll l_{Y,c}$ represents the mesopores. The void phase is defined as the union of their excursion sets,
\begin{equation}
    I_{\mathrm{void}}(\mathbf{x}) = H[a_f - Y_f(\mathbf{x})] \;\vee\; H[a_c - Y_c(\mathbf{x})].
    \label{eq:twofield}
\end{equation}
The solid is the intersection of the two solid phases, so that the mesopores form a population independent of the macropores rather than a roughening of a single interface (Fig.~\ref{fig:grf_xsections}). By independence, the porosity of the union is
\begin{equation}
    \phi_0 = 1 - (1-\phi_0^{(f)})(1-\phi_0^{(c)}) = \phi_0^{(f)} + \phi_0^{(c)} - \phi_0^{(f)}\,\phi_0^{(c)},
    \label{eq:unionporosity}
\end{equation}
where $\phi_0^{(f)}$ and $\phi_0^{(c)}$ are the void fractions that each field would produce alone. Since the surface area of a level-cut field scales as $1/l_Y$, essentially all of the surface area is contributed by the fine field (Appendix~\ref{app:grf}). The two scales therefore have separate roles. The fine field sets the surface area and the mesopore volume, and the coarse field the size of the flow-carrying macropores. This two-field construction is the model proposed in this work and is referred to as the proposed model hereafter.

The fine-field parameters $l_{Y,f}$ and $\phi_0^{(f)}$ are calibrated to the pore volume and surface area resolved by nitrogen sorption within the BJH window ($1.9$ to $164$~nm), and the coarse threshold then follows from the density-derived porosity through Eq.~\eqref{eq:unionporosity}. The coarse-field correlation length cannot be calibrated in the same way, since no sorption measurement constrains the macropore size. It is bounded below by the requirement that the macropore population lie above the BJH window, and above by the requirement that the simulation domain contain several coarse-field correlation lengths.

The upper bound on $l_{Y,c}$ comes from the electron micrograph (Fig.~\ref{fig:sem}), to which the reconstructions are adapted through the mean chord length of the void regions. The measured chords of $80$ to $170$~nm are reproduced for $l_{Y,c}$ between $100$ and $150$~nm, whereas larger correlation lengths give systematically longer chords. The simulation boxes are sized relative to this bound, with an edge length of about $17$~$l_{Y,c}$. As both samples share the same synthesis route, we use $l_{Y,c} = 150$~nm for each of them, with an uncertainty of about $30\%$ that is independent of the permeability analysis.

\FloatBarrier
\subsection{\label{sec:comp:structures}Microstructure ensembles}
Two ensembles of single-field microstructures were generated to examine the relationship between morphology and gas permeability. The order parameter was fixed at $\nu = 1$, which produces moderately disordered morphologies. The influence of $\nu$ on pore size is examined in Appendix~\ref{sec:comp:nueffect}. The remaining free parameters, porosity $\phi_0$ and correlation length $l_Y$, were varied over the range typical of supercritically dried RF aerogels. Six porosities ($\phi_0 = 0.70$ to $0.95$ in increments of $0.05$) were combined with three correlation lengths ($l_Y = 20$, $40$, and $60$~nm). Each of the 18 parameter combinations was realized using three independent random seeds, resulting in a total of 54 microstructures. Comparison with the experimental samples required a second family of structures at the experimental porosity. A single-field GRF set was therefore generated at the fixed porosity $\phi_0 = 0.906$, corresponding to the measured samples. The correlation length was varied over $l_Y = 10$, 15, 20, 50, 100, 200, and 300~nm, with three realizations per correlation length, giving 21 structures in total. The reconstructions with the proposed model of Sec.~\ref{sec:comp:grf} were generated for both samples at $l_{Y,c} = 150$~nm with three realizations each, and for a sweep of the coarse-field correlation length over $l_{Y,c} = 100$, $150$, $200$, $300$ and $450$~nm at fixed fine-field calibration. The domain size and voxel resolution used for these ensembles were selected by the numerical convergence analysis and are presented in detail in Sec.~\ref{sec:comp:convergence}. At the experimental porosity, a single level-cut field does not form a connected solid, whereas every reconstruction with the proposed model does. The connectivity analysis and the effect of the order parameter on pore size are given in Appendix~\ref{app:grf}.

\FloatBarrier
\section{\label{sec:comp:perm}\label{sec:results}Permeability models and transport results}

\subsection{\label{sec:comp:analytical}Analytical permeability models}

The permeability of a porous medium quantifies its ability to transmit fluid under a pressure gradient. For continuum (Darcy) flow of a viscous fluid through an isotropic porous medium, the most widely used analytical estimate is the Kozeny-Carman (KC) equation~\citep{kozeny1927, carman1937}:
\begin{equation}
    k = \frac{\phi_0^3}{k_0 \, \tau^2 \, S_0^2 \, (1 - \phi_0)^2},
    \label{eq:kc}
\end{equation}
where $\phi_0$ is the porosity, $S_0$ is the specific surface area per unit solid volume, $\tau$ is the geometric tortuosity defined as the ratio of the mean flow path length to the straight-line distance through the medium, and $k_0$ is a shape factor determined by the cross-sectional geometry of the pore channels. The product $c = k_0 \tau^2$ is often referred to as the Kozeny constant. The classical value of approximately $c = 5$ was obtained empirically for packed beds of uniform spheres with $k_0 \approx 2.5$ and $\tau$ close to $\sqrt{2}$~\citep{carman1937}. Equation~\eqref{eq:kc} is based on the assumption that (i) the pore space can be represented as a bundle of tortuous, non-intersecting capillaries of uniform cross-section, (ii) the flow is laminar and governed by the no-slip condition at the pore walls, and (iii) inertial effects are negligible (Stokes regime). These assumptions are well satisfied in macroporous granular media, where the pore sizes are large compared to the molecular mean free path and the pore geometry is relatively regular.

However, for micro-/meso-porous materials such as organic aerogels, several of these assumptions break down simultaneously. First, aerogel microstructures exhibit complex, bi-continuous morphologies with broad pore size distributions and interconnected pore geometries that cannot be adequately described by a single hydraulic radius. Second, the use of a constant Kozeny factor $c$ presumes that the tortuosity and shape factor are independent of porosity and pore morphology, an assumption that becomes increasingly inaccurate for highly porous, disordered structures ($\phi_0 > 0.7$)~\citep{scherer1994,xu2008}. A further and more fundamental limitation arises from gas rarefaction effects. In aerogels, pore diameters typically lie in the mesoporous range ($d$ roughly $2$-$50$~nm), which is comparable to the mean free path of common gases at ambient conditions. The flow regime is characterized by the Knudsen number $\mathrm{Kn} = \lambda/l_c$, where $l_c$ is a representative length of the pore space. Throughout this work we define the characteristic length in terms of the throat radius, $\mathrm{Kn} = \lambda/R$. This is based on the convention of the Beskok-Karniadakis correlation \citep{beskok1999}. At the sample scale the natural choice of $l_c$ is the mean chord length of the pore phase, $4\phi_0/S$ \citep{Levitz:1993}, which is approximately 75 to 80\,nm for the present samples. The Knudsen number thus reaches values of order unity and places the transport in the transition regime, where the no-slip boundary condition assumed in the KC derivation is no longer valid.

\citet{klinkenberg1941} proposed a first-order correction to account for slip flow, expressing the apparent gas permeability as
\begin{equation}
    k_{\mathrm{app}} = k_\infty \left(1 + \frac{b_{\mathrm{K}}}{p}\right),
    \label{eq:klinkenberg}
\end{equation}
\noindent where $k_\infty$ is the intrinsic permeability, $b_{\mathrm{K}}$ is the Klinkenberg slip parameter (dependent on the gas species, temperature, and mean pore size), and $p$ is the mean gas pressure. While Eq.~\eqref{eq:klinkenberg} provides a useful empirical correction in the slip-flow regime ($\mathrm{Kn} \lesssim 0.1$), it becomes insufficient in the transition regime ($0.1 < \mathrm{Kn} < 10$) where the linear pressure dependence no longer holds.

A more general treatment applicable across all flow regimes was developed by \citet{beskok1999}, who proposed a unified model for the volumetric flow rate in channels and ducts that smoothly interpolates between the continuum, slip, transition, and free-molecular regimes. The model introduces a Knudsen-dependent correction to the Hagen-Poiseuille conductance:
\begin{equation}
    g^{\mathrm{BK}} = g^{\mathrm{HP}}
    \left(1 + \alpha\,\mathrm{Kn}\right)
    \left(1 + \frac{4\,\mathrm{Kn}}{1 - b\,\mathrm{Kn}}\right),
    \label{eq:bk}
\end{equation}
where $\alpha = \alpha_0\,(2/\pi)\,\arctan(\alpha_1\,\mathrm{Kn}^\beta)$ is a rarefaction coefficient with parameters $\alpha_0 = 64/(15\pi)$ (approximately $1.358$), $\alpha_1 = 4.0$, $\beta = 0.4$, and $b = -1$ is the slip coefficient. We denote the resulting correction factor by
\begin{equation}
C_r(\mathrm{Kn}) = (1 + \alpha\,\mathrm{Kn})\left(1 + 4\,\mathrm{Kn}/(1 - b\,\mathrm{Kn})\right),
\end{equation}
which satisfies $C_r \to 1$ in the continuum limit with $\mathrm{Kn} \to 0$ and grows linearly with $\mathrm{Kn}$ in the free-molecular limit.

Given the complex pore morphology, high porosity, and strong rarefaction effects present in mesoporous aerogels, analytical models alone are insufficient for reliable permeability prediction. This motivates the use of pore-network modelling, in which the microstructure is resolved explicitly and the BK model is applied at the throat level to capture the local flow physics across all Knudsen regimes.

\subsection{\label{sec:comp:pnm}Pore-network modelling of gas flow}

\begin{figure*}
    \centering
    \begin{subfigure}[b]{0.25\linewidth}
        \centering
        \includegraphics[width=\linewidth]{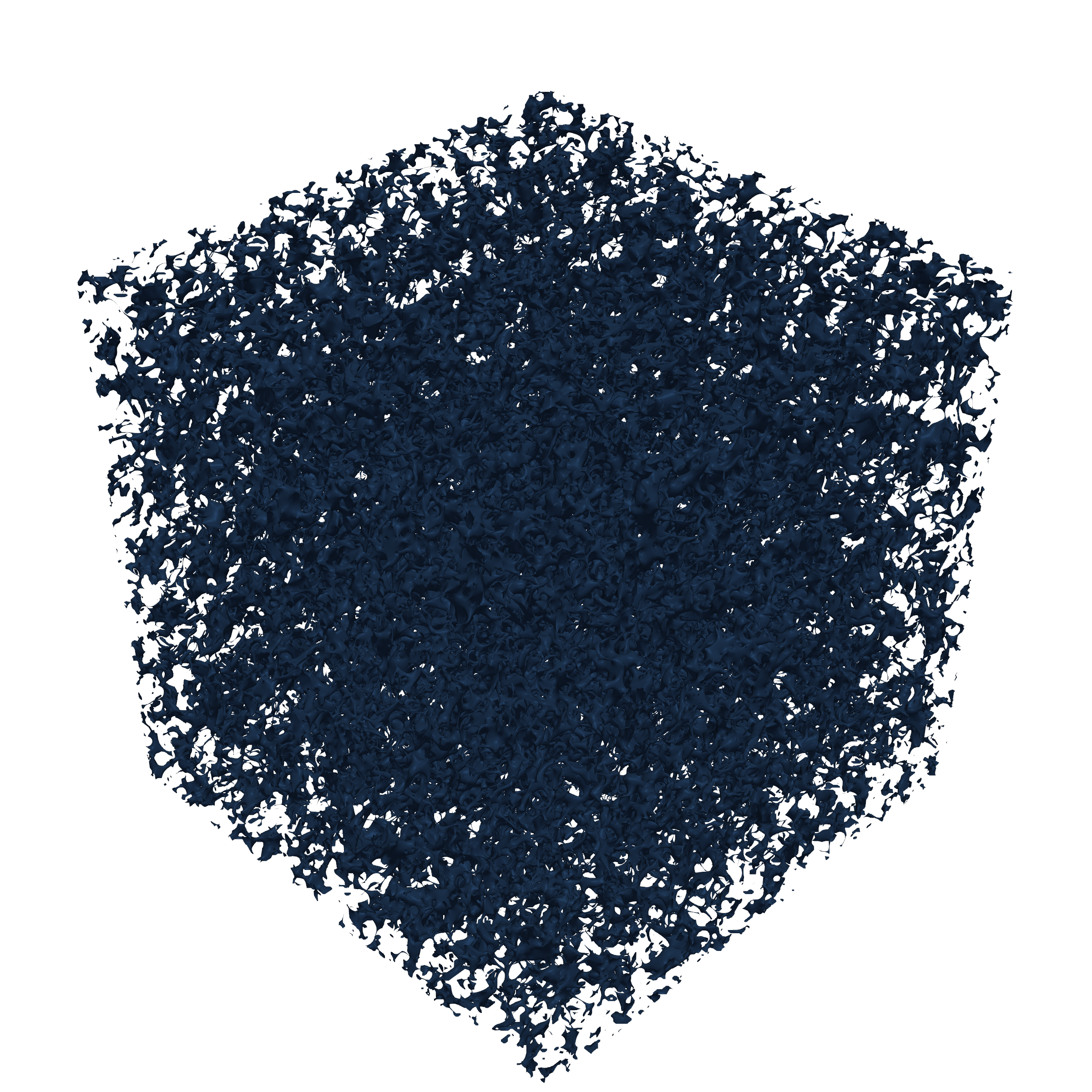}
        \caption{}
        \label{fig:surface_3d}
    \end{subfigure}
    \begin{subfigure}[b]{0.3\linewidth}
        \centering
        \includegraphics[width=\linewidth]{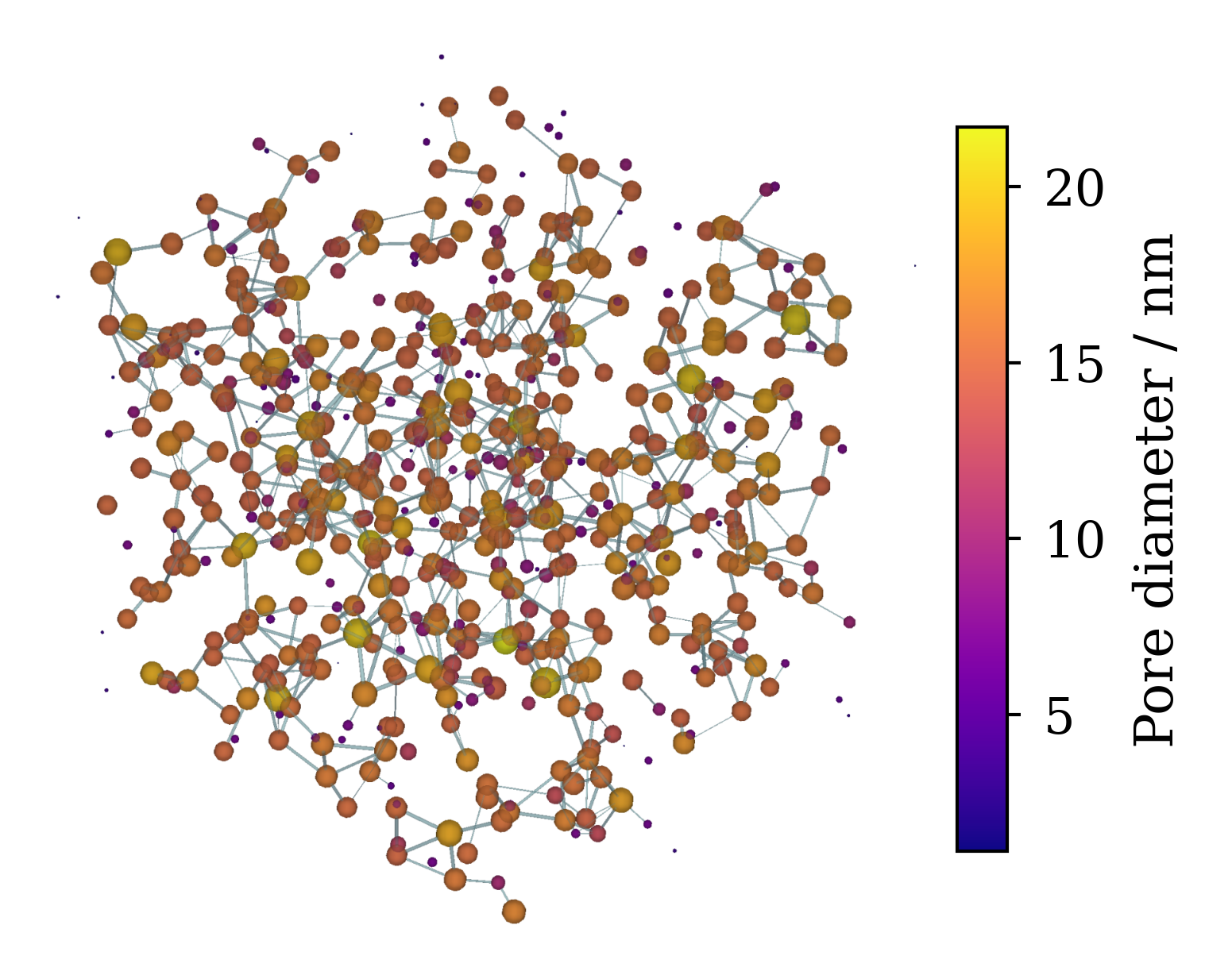}
        \caption{}
        \label{fig:pore_network}
    \end{subfigure}
    \hspace{0.02\linewidth}
    \begin{subfigure}[b]{0.38\linewidth}
        \centering
        \includegraphics[width=\linewidth]{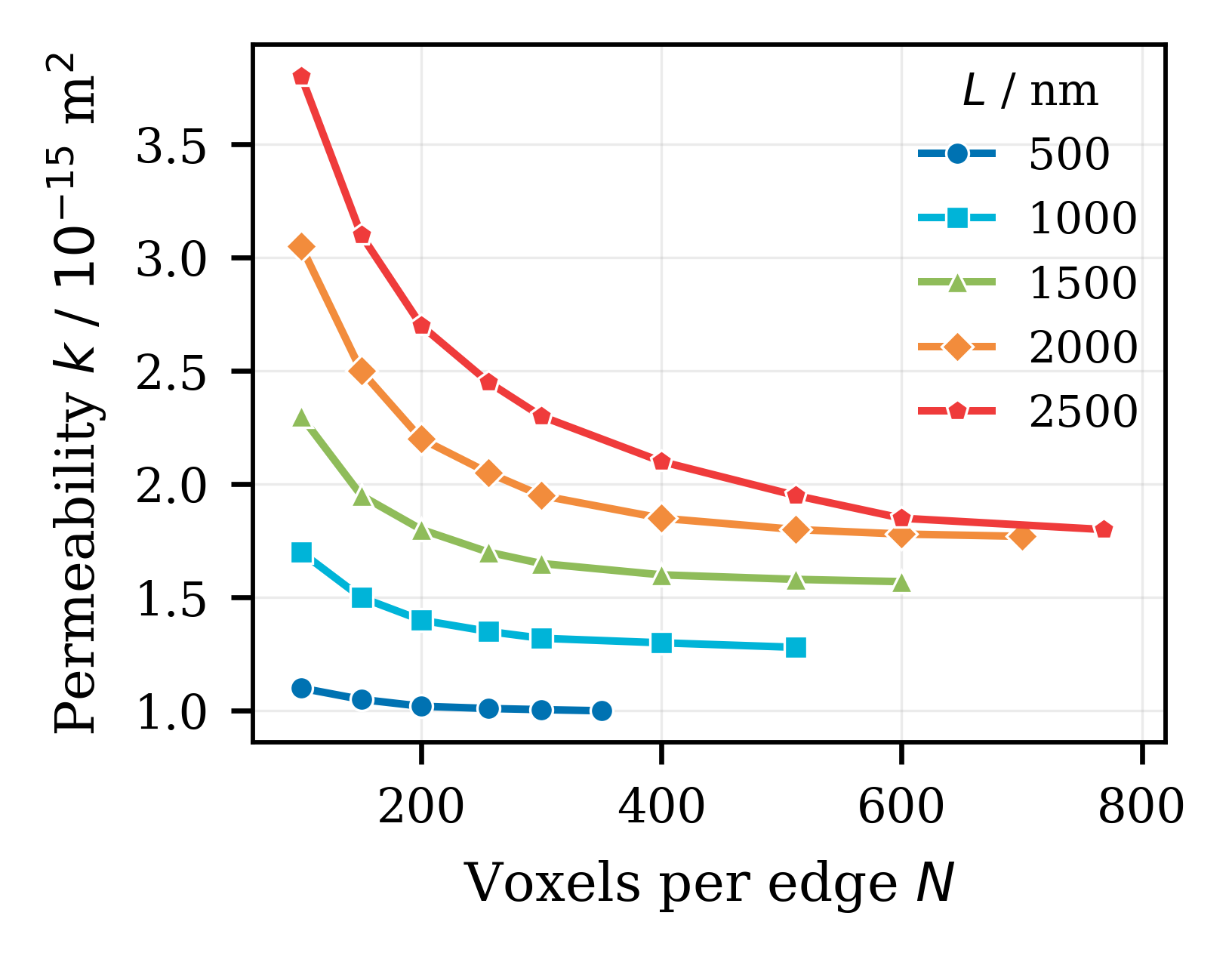}
        \caption{}
        \label{fig:permeability_resolution}
    \end{subfigure}

    \caption{(a)~Surface rendering of a representative GRF microstructure. (b)~Extracted pore network (pores scaled and coloured by diameter). (c)~Convergence of the computed permeability with voxel resolution for increasing domain edge $L$.}
    \label{fig:microstructure_and_convergence}
\end{figure*}

\paragraph{Network extraction}
The pore network was extracted from the binary microstructure using the SNOW algorithm~\citep{gostick2017versatile}, in its SNOW2 implementation in PoreSpy~\citep{gostick2019porespy}. The algorithm operates on the Euclidean distance transform $D(\mathbf{x})$ of the pore phase to identify local maxima, which serve as markers for a watershed segmentation. Adjacent basins sharing saddle points below a merging threshold are combined to eliminate over-segmentation, yielding a graph $\mathcal{G} = (\mathcal{V}, \mathcal{E})$ where nodes $\mathcal{V}$ represent pores characterized by equivalent diameter $d_i$ and volume $V_i$, and edges $\mathcal{E}$ represent throats with length $L_{ij}$ and cross-sectional area $A_{ij}$. Two size measures follow from this construction and are used throughout. The pore diameter $d_i$ is the diameter of the sphere having the same volume as watershed basin $i$, while the throat diameter $d_{t,ij} = 2R_{t,ij}$ is set by the inscribed radius at the constriction, $R_{t,ij} = \max D(\mathbf{x})$ over the surface separating basins $i$ and $j$. The throat diameter is therefore a constriction size and is necessarily smaller than the diameters of the two pores it joins. A representative microstructure constructed with single-field GRF and its extracted network, with pore bodies scaled and coloured by their equivalent diameter and connected by semi-transparent throats, are shown in Figs.~\ref{fig:surface_3d} and \ref{fig:pore_network}.

Single-phase gas flow through the extracted network was simulated in OpenPNM~\citep{gostick2016openpnm}. Each throat is treated as a cylindrical conduit with the Hagen-Poiseuille conductance $\pi R^4/(8\mu L)$~\citep{Bear1972,Dullien1992}, with the inscribed throat radius $R_t$ from the SNOW2 extraction as $R$ and the pore centre-to-centre distance $L_{ij}$ as $L$, rescaled by the rarefaction factor of Eq.~\eqref{eq:bk}: 
\begin{equation}
    g_{ij} = \frac{\pi\,R_{t}^{4}}{8\,\mu\,L_{ij}}\,C_r(\mathrm{Kn}_{ij}),
    \label{eq:throat_conductance}
\end{equation}
where $C_r$ is evaluated at the local Knudsen number $\mathrm{Kn}_{ij} = \lambda / R_t$, based on the throat radius, with the mean free path $\lambda$ taken at the arithmetic mean pressure of the two neighbouring pores. A throat-only conductance formulation is used, in which the BK correction is applied exclusively at the throat constriction and the full centre-to-centre distance serves as the effective hydrodynamic path length. This avoids applying the cylindrical-tube rarefaction model to the large, multiply connected pore-body cavities where its assumptions are not satisfied, an issue that becomes acute at porosities above 0.85 where the SNOW2-extracted pore bodies overlap extensively. A detailed comparison of this formulation against two alternative conduit models is provided in Appendix~\ref{app:conductance}. The pore pressures follow from mass conservation. At steady state the net flow into every interior pore vanishes,
\begin{equation}
    \sum_{j \in \mathcal{N}(i)} g_{ij}\,(P_i - P_j) = 0,
    \label{eq:massbalance}
\end{equation}
where $\mathcal{N}(i)$ is the set of pores connected to pore $i$. Equation~\eqref{eq:massbalance} would be a sparse linear system in the $P_i$, but the conductances depend on the pressures themselves, since the mean free path $\lambda = k_B T / (\sqrt{2}\,\pi\,d_{\mathrm{mol}}^2\,P)$ entering $\mathrm{Kn}_{ij}$ is evaluated at the local pressure. This weak nonlinearity was resolved by Picard iteration, in which the conductances are frozen at the previous pressure iterate and the resulting linear system is solved,
\begin{equation}
    \sum_{j \in \mathcal{N}(i)} g_{ij}\big(P^{(k)}\big)\,\big(P_i^{(k+1)} - P_j^{(k+1)}\big) = 0 .
    \label{eq:picard}
\end{equation}
The iteration starts from a linear pressure profile between inlet and outlet. At each step, $\lambda$, $\mathrm{Kn}_{ij}$ and the conductances of Eq.~\eqref{eq:throat_conductance} are recomputed from the current pressures, and Eq.~\eqref{eq:picard} is solved again until the relative change in permeability falls below $10^{-4}$. A pressure difference $\Delta P = P_{\mathrm{in}} - P_{\mathrm{out}}$ is imposed between opposite faces of the cubic domain, and the remaining faces are treated as no-flow walls. The simulations reproduce the experimental conditions, with $T = 293$~K, $\mu = 1.81\cdot10^{-5}$~Pa\,s, $d_{\mathrm{mol}} = 3.7\cdot10^{-10}$~m and $P_{\mathrm{out}} = 101\,325$~Pa. The measured chamber pressures lie between $940$ and $1034$~mbar, so simulation and experiment operate at comparable mean free paths and Knudsen numbers. Network models resolve the connectivity and constriction sizes of the pore space but approximate the flow within each conduit, and the choice of throat conductance alone changes the absolute permeability by up to a factor of $2.6$ (Appendix~\ref{app:conductance}). The pore and throat size distributions are further sensitive to the length scale of the watershed segmentation that partitions the pore space, which introduces a variation of approximately $\pm 25\%$. Absolute permeabilities are therefore reliable to within a factor of two to three, whereas the conclusions below rest on differences of one to two orders of magnitude.

The bulk specific surface area $S_{\mathrm{bulk}}$ was computed directly from the binary microstructure using the marching cubes algorithm~\citep{lorensen1987} to extract the solid-void isosurface at a level set of 0.5. The total interfacial area was obtained by summing the triangle areas of the resulting mesh, yielding $S_{\mathrm{bulk}}$ in units of m$^2$\,m$^{-3}$. The hydraulic tortuosity is evaluated from the converged flow field as the flux-weighted ratio of travelled length to axial advance over all throats~\citep{Koponen1996},
\begin{equation}
    \tau_h = \frac{\sum_{ij} |Q_{ij}|\,L_{ij}}{\sum_{ij} |Q_{ij}|\,|\Delta a_{ij}|},
    \label{eq:tortuosity}
\end{equation}
where $L_{ij}$ is the pore centre-to-centre distance of throat $ij$, $|\Delta a_{ij}|$ its projection on the macroscopic flow direction, and $Q_{ij}$ its volumetric flow rate. Its isotropic limit is discussed in Sec.~\ref{sec:results:tortuosity}.  These quantities were used to evaluate Eq.~\eqref{eq:kc} with $S_0 = S_{\mathrm{bulk}}/(1-\phi_0)$ and $k_0 = 2$. All quantities except $k_0$ are computed directly from the microstructure or its flow solution without free parameters.

\subsection{\label{sec:comp:convergence}Numerical convergence and simulation-domain selection}

To establish numerical convergence of the permeability, a sensitivity analysis was performed with respect to computational domain size and spatial resolution. The domain edge length ($L=500$--$2500$~nm) and the number of voxels per edge ($N=100$--$768$) were systematically varied. As shown in Fig.~\ref{fig:permeability_resolution}, the computed permeability depends strongly on resolution at coarse discretizations but approaches a stable value upon refinement. Larger domains require more voxels per edge to achieve convergence, since the characteristic microstructural length scales must remain adequately resolved. Based on these results, $L=2500$~nm and $N=768$, corresponding to a voxel size of $h=3.26$~nm, were adopted for subsequent simulations.

The reconstructions with the proposed model require a separate resolution assessment because the fine-field correlation length is represented by only approximately 1.3 voxels at this resolution. A 7A realization was therefore generated at twice the spatial resolution ($h=1.63$~nm) in a $1250$~nm domain, the largest domain accommodated by 768 voxels per edge. A corresponding $3.26$~nm representation was obtained by block averaging and re-imposing the prescribed porosity, providing two discretizations of the same underlying structure (Table~\ref{tab:twofield_convergence}). The flux-weighted throat diameter, the corresponding flux-weighted Knudsen number and the enhancement factor $k_{\mathrm{app}}/k_\infty$ differ by less than $2\%$ between the two resolutions. The absolute permeability, however, is approximately $25\%$ higher on the finer grid. This difference originates primarily from network extraction rather than from the flow calculation. The watershed algorithm merges peaks of the distance map within a radius specified in voxels; consequently, halving the voxel size while retaining the same merging radius reduces its physical extent from approximately 13 to 6.5~nm. This leads to finer partitioning of the macropores and approximately eight times as many extracted throats. When the merging radius is instead increased to eight voxels to retain its physical value of 13~nm, the resulting permeability is 16--22\% lower than that obtained on the $3.26$~nm grid. We therefore assign a conservative discretization uncertainty of $\pm25\%$ to the absolute permeability, associated with the resolution sensitivity of network extraction. This uncertainty is additional to the conductance-formulation uncertainty discussed in Appendix~\ref{app:conductance}. The influence of domain size is smaller: separate tests over $L=625$--$2500$~nm show permeability variations of less than $\pm9\%$.

\begin{table}
\begin{center}
\begin{tabular}{lccc}
\hline
Voxel size $h$ (nm) & 3.26 & 1.63 & 1.63$^{\dagger}$ \\
$k$ ($10^{-15}$~m$^2$) & 6.0 & 7.5 & 4.7 \\
$k_{\mathrm{app}}/k_\infty$ & 3.49 & 3.53 & 3.53 \\
Flux-weighted throat diameter (nm) & 310 & 307 & 304 \\
$S_{\mathrm{bulk}}$ ($10^{7}$~m$^2$\,m$^{-3}$) & 3.4 & 4.0 & 4.0 \\
Throats ($10^{3}$) & 34 & 267 & 68 \\
\hline
\end{tabular}
\caption{Grid test of one 7A realization of the proposed 
model ($l_{Y,c} = 150$~nm, box edge $1250$~nm, 
$\Delta P = 50$~mbar) at $h = 3.26$ and $1.63$~nm. 
$^{\dagger}$Watershed peak-merging radius held at 
$13$~nm instead of $4$ voxels.}
\label{tab:twofield_convergence}
\end{center}
\end{table}

\FloatBarrier
\subsection{\label{sec:results:transport}Gas transport and apparent permeability}

With the domain and grid fixed, the ensembles of Sec.~\ref{sec:comp:structures} were run at the conditions of Sec.~\ref{sec:comp:pnm}. Because the mean free path of air at atmospheric pressure and 293~K is approximately 66~nm, which is comparable to the throat diameters, all simulations operate within the slip-flow and transition regimes. Across the parametric grid, the mean Knudsen number at $\Delta P = 50$~mbar ranges from 1.9 for the structures with the widest throats to 6.3 for those with the narrowest. The proposed model reaches a mean Knudsen number of 17 (7A at $l_{Y,c} = 150$~nm, averaged over three structures), exceeding that of all single-field structures considered here because its throat population is dominated by mesopores formed within the solid aggregates. This heterogeneity necessitates a throat-scale rarefaction model, since applying a single bulk Knudsen correction uniformly throughout the network would underestimate slip in the narrowest throats while overestimating it in the widest. Because the Hagen-Poiseuille conductance scales with the fourth power of diameter, the throats rather than the pore bodies constitute the hydraulic bottlenecks of all networks.

The apparent gas permeability $K_{\mathrm{app}} = \tfrac{1}{3}(K_{xx}+K_{yy}+K_{zz})$ of the parametric structures spans $2.5 \cdot 10^{-16}$ to $5.0 \cdot 10^{-15}$~m$^2$ ($\phi_0 = 0.70$-$0.95$, $l_Y = 20$-$60$~nm), consistent with the lower end of permeabilities reported for carbon aerogels~\citep{Kong1993} and above those of silica aerogels~\citep{Woignier2023}, and the directional anisotropy $K_{\max}/K_{\min} = 1.07 \pm 0.04$ confirms macroscopic isotropy, as expected for an isotropic random field. Because the local Knudsen numbers extend into the transition regime, slip flow significantly enhances gas transport. Consequently, the apparent permeability decreases by an average of 10\% over the pressure range $\Delta P = 25$-$300$~mbar. The intrinsic permeability $K_\infty$ is obtained directly from the same networks by setting $C_r = 1$ in every throat. The ratio $K_{\mathrm{app}}/K_\infty$ at $\Delta P = 50$~mbar ranges from $5.6$ for the most open structures ($\phi_0 = 0.95$, $l_Y = 60$~nm) to $27$ for the densest ($\phi_0 = 0.70$, $l_Y = 20$~nm). A linear Klinkenberg extrapolation (Eq.~\eqref{eq:klinkenberg}) over the simulated pressure range, which covers mean pressures of only $1.03$ to $1.16$~bar, would give $4.3$ to $8.9$ and thus underestimate the enhancement by up to a factor of three, as expected at mean Knudsen numbers between $2$ and $6$, where the pressure dependence is no longer linear.

The Kozeny-Carman equation (Eq.~\eqref{eq:kc}), evaluated using the porosity, specific surface area, and hydraulic tortuosity extracted from each microstructure, has a substantially lower predictive accuracy. It underpredicts the permeability of the densest structures by up to a factor of 6.1, owing to the omission of slip-flow effects, while overpredicting the most porous structures ($\phi_0 = 0.95$) by up to a factor of $2.2$, where the hydraulic-radius approximation breaks down, and transport becomes increasingly governed by fibre-drag mechanisms~\citep{Happel1965,duPlessis1994}. The geometric-mean ratio of network to Kozeny-Carman permeability is $2.0$, and the coefficient of determination of the logarithmic parity is negative ($R^2 = -0.59$). Parametric maps, including directional permeabilities, pressure dependence, Kozeny-Carman parity and residual plots, and permeability-porosity scaling, are provided in Appendix~\ref{app:paramaps}.

\subsubsection{\label{sec:results:envelope}Dependence on surface area and porosity}

\begin{figure*}
\centering
\includegraphics[width=0.85\textwidth]{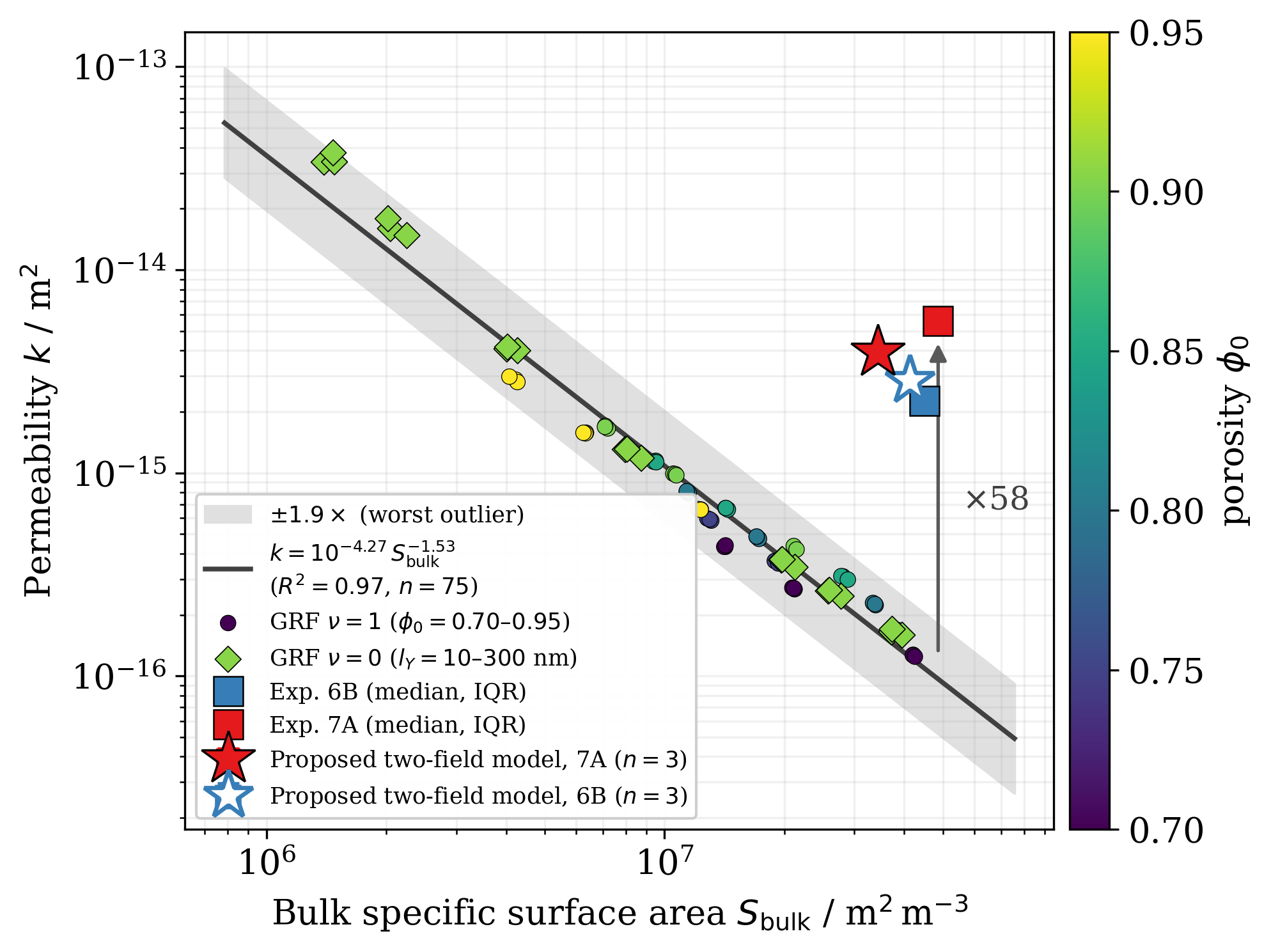}
\caption{\label{fig:master_curve}
Apparent permeability against bulk specific surface area for all 75 single-field structures ($\nu=1$ grid, circles; $\nu=0$ fixed-porosity set, diamonds), together with the pooled fit (line, Eq.~\eqref{eq:master}) and worst-outlier band (shading). The measured samples (squares) and the proposed model (stars) lie above the single-field envelope.}
\end{figure*}

The Kozeny-Carman relation, Eq.~\eqref{eq:kc}, predicts an inverse-square dependence of permeability on specific surface area at fixed porosity. We examine this dependence using 75 single-field structures, comprising 54 realizations with $\nu=1$, $\phi_0=0.70$-$0.95$ and $l_Y=20$-$60$~nm, and 21 realizations with $\nu=0$, $\phi_0=0.906$ and $l_Y=10$-$300$~nm. Despite the variation in porosity, correlation length and order parameter, the apparent permeabilities collapse onto a common trend when plotted against the bulk specific surface area, $S_\mathrm{bulk}$ (Fig.~\ref{fig:master_curve}). For air at 293~K and atmospheric pressure, a least-squares fit in logarithmic space gives
\begin{equation}
\label{eq:master}
K=10^{-4.80}S_\mathrm{bulk}^{-1.42},
\qquad R^2=0.97,
\qquad n=75,
\end{equation}
where $K$ is in m$^2$ and $S_\mathrm{bulk}$ in m$^2$\,m$^{-3}$. Within the investigated parameter space, bulk specific surface area provides a strong predictor of apparent permeability, although it does not uniquely determine it, with no structure deviating by more than a factor of $1.68$.

The fitted exponent of $-1.42$ is weaker than the inverse-square scaling expected for geometrically similar structures at fixed porosity. In the no-slip limit, the characteristic length scales as $S_\mathrm{bulk}^{-1}$ and hence $K_\infty\propto S_\mathrm{bulk}^{-2}$, consistent with the pooled no-slip and Kozeny-Carman exponents of $-2.21$ and $-2.18$, respectively. The weaker apparent scaling results from rarefaction: for a fixed gas and pressure, the permeability can be expressed as
\begin{equation}
K=l_Y^2 f\left(\phi_0,\frac{\lambda}{l_Y}\right),
\end{equation}
where $\lambda$ is the molecular mean free path. Reducing the characteristic pore size increases the Knudsen number and the relative contribution of slip, thereby weakening the dependence of permeability on surface area. This is confirmed by the fixed-porosity series, for which the local surface-area exponent varies from $-1.21$ to $-1.89$ as $l_Y$ increases from 10 to 300~nm. The underlying structures and detailed scaling exponents are reported in Appendix~\ref{app:locus} and Appendix~\ref{app:scaling}, respectively, with the full exponent comparison given in Table~\ref{tab:scaling_summary}.

The residual dependence on porosity is comparatively weak once the bulk specific surface area is accounted for. A joint fit gives
\begin{equation}
K\propto S_\mathrm{bulk}^{-1.38}\phi_0^{0.61},
\end{equation}
corresponding to a permeability variation of only a factor of approximately 1.2 across $\phi_0=0.70$-$0.95$ at fixed surface area. The influence of the order parameter is similarly limited: at comparable porosity and $l_Y=20$~nm, structures with $\nu=0$ and $\nu=1$ differ in permeability by approximately 20\%. These comparisons show that surface area captures most of the variation across the investigated structures, with porosity and the order parameter providing secondary corrections (Appendix~\ref{app:scaling}).

\begin{table}
\begin{center}
\begin{tabular}{lccc}
\hline
Scaling variable & $m_{\mathrm{app}}$ & $m_{\infty}$ & $m_{\mathrm{KC}}$ \\
\hline
$\bar{d}_{\mathrm{pore}}$ (fixed $\phi_0$)
& 1.34-1.57 & 2.56-2.76 & 2.49-2.54 \\
$\bar{d}_{\mathrm{thr}}$ (fixed $\phi_0$)
& 1.10-1.27 & 2.12-2.23 & 2.04-2.07 \\
$\phi_0$ (fixed $l_Y$)
& 5.25-5.81 & 6.84-7.15 & 10.50-10.70 \\
$S_{\mathrm{bulk}}$ (all 75 structures)
& $-1.42$ & $-2.21$ & $-2.18$ \\
$S_{\mathrm{bulk}}$ ($\phi_0=0.906$)
& $-1.21$ to $-1.89$ & $-2.2$ to $-2.6$ & $-2.0$ to $-2.1$ \\
\hline
\end{tabular}
\caption{Power-law exponents $m$ for $K\propto x^m$. 
The columns compare the apparent permeability $K$, 
the no-slip permeability $K_\infty$ of the same networks 
($C_r=1$ in every throat), and the Kozeny-Carman estimate 
$K_{\mathrm{KC}}$. All fits have $R^2\geq0.97$. 
Exponents at $\phi_0=0.906$ are local slopes between 
neighbouring values of $l_Y$.}
\label{tab:scaling_summary}
\end{center}
\end{table}

\subsubsection{\label{sec:results:tortuosity} Hydraulic and diffusive tortuosity}

The flux-weighted hydraulic tortuosity of Equation~\eqref{eq:tortuosity} is nearly constant across every structure simulated in this work. Its mean value is $\tau_h = 1.516$ with a full range of 1.484 to 1.541 over all 25 single-field structure groups (a group being the seed average of one parameter combination), spanning porosities from 0.70 to 0.95, mean pore sizes from 43 to 267~nm and both values of the order parameter $\nu$ (18 groups at $\nu = 1$ averaging $1.513$ and seven at $\nu = 0$ averaging $1.523$), with the directional components of each group agreeing to within 4 per cent. The reconstructions with the proposed model lie between 1.52 and 1.57 over the admissible macropore scale $l_{Y,c} = 100$-$150$~nm (three-seed mean $1.54$ for 7A at $150$~nm), and the individual realizations of the full coarse-field sweep of Sec.~\ref{sec:comp:structures}, $l_{Y,c} = 100$-$450$~nm, between 1.46 and 1.62, so the hierarchy leaves the hydraulic tortuosity essentially unchanged. This near-constancy reflects the estimator rather than the material. For an affine pressure field $p = -Gz$, a throat of length $L_{ij}$ and conductance $g_{ij}$ inclined at $\theta_{ij}$ to the flow axis carries $|Q_{ij}| = g_{ij} G L_{ij} |\cos\theta_{ij}|$ and advances the gas by $|\Delta a_{ij}| = L_{ij}|\cos\theta_{ij}|$. Equation~\eqref{eq:tortuosity} then reduces to
\begin{equation}
    \tau_h = \frac{\sum_{ij} g_{ij} L_{ij}^2 |\cos\theta_{ij}|}{\sum_{ij} g_{ij} L_{ij}^2 \cos^2\theta_{ij}} \;\to\; \frac{\langle|\cos\theta|\rangle}{\langle\cos^2\theta\rangle} = \frac{3}{2}
    \label{eq:tauh_affine}
\end{equation}
for any network whose throat orientations are isotropic and uncorrelated with $g_{ij}$ and $L_{ij}$. The computed values lie within $5\%$ of this limit, so $\tau_h$ is not a transferable material constant, and only its small deviations from $3/2$, which measure the departure of the pressure field from affinity, carry information on the flow paths.

The diffusive tortuosity $\tau_{fd}$, obtained from a voxel-level finite-difference solution of the Laplace problem on the void phase (computational details in Appendix~\ref{app:taufd}), behaves differently and separates the two constructions. For every single-field structure, from correlation lengths of 20 to 300~nm, it lies between 1.07 and 1.09. The structures of the proposed model remain at the isotropic limit of $\tau_h$, while their diffusive tortuosity (Table~\ref{tab:app_taufd}) rises to $1.14 \pm 0.01$ for 7A and $1.19 \pm 0.05$ for 6B at $l_{Y,c} = 150$~nm, and to $1.14$ and $1.15$ at $l_{Y,c} = 300$~nm. Unlike $\tau_h$, the diffusive tortuosity follows from a full solution of the Laplace problem on the voxel geometry and is not tied to an orientational limit. Diffusive transport samples the mesopore space in proportion to its volume, so that $\tau_{fd}$ increases with the mesopore share of the pore volume, as between 7A and 6B. The diffusive tortuosity therefore serves as a diagnostic of hierarchy, since single-field media sit at $\tau_{fd} \approx 1.08$, whereas a diffusive excess indicates a multi-scale pore space. The two quantities are distinct observables and should not be conflated when values reported in the literature are compared.

\FloatBarrier
\subsection{\label{sec:results:validation}Comparison with experimentally measured permeabilities}

Samples 6B and 7A provide a useful test of whether the reconstructed pore structure captures the features governing gas transport. As discussed in Sec.~\ref{sec:exp}, the two samples have comparable porosities ($\phi_0 \approx 0.905$) and BET specific surface areas ($318$ and $345$~m$^2$\,g$^{-1}$), yet their measured permeabilities differ by a factor of approximately $2.5$, with values of $2.26$ and $5.62\cdot10^{-15}$~m$^2$, respectively (Fig.~\ref{fig:exp_permeability}). This difference suggests that porosity and total surface area alone are insufficient to characterize the transport properties, and that the spatial organization of the pore space plays an essential role.

At the experimental porosity, the single-field permeability envelope of Sec.~\ref{sec:results:envelope} falls between factors of $10$ and $29$ below the measured values (Appendix~\ref{app:locus}). This indicates that a single characteristic length scale cannot simultaneously capture the fine structure responsible for the measured surface area and the larger-scale connectivity required for hydraulic transport.

The proposed model addresses this limitation by introducing a coarse structural scale while retaining the fine field constrained by sorption measurements (Sec.~\ref{sec:comp:grf}). With the coarse threshold determined from the macroporosity and $l_{Y,c}=150$~nm estimated from microscopy, the predicted permeabilities are $(6.1\pm0.3)\cdot10^{-15}$~m$^2$ for 7A and $(4.5\pm0.3)\cdot10^{-15}$~m$^2$ for 6B. The corresponding ratios of predicted to measured permeability are 1.1 and 2.0, respectively. Combining the $\pm5\%$ realization scatter with the $\pm25\%$ discretization uncertainty described in Sec.~\ref{sec:comp:convergence} gives ratios of $1.1\pm0.3$ and $2.0\pm0.5$. Thus, the agreement for 7A is consistent with the quantified uncertainty but does not independently validate the model. For 6B, the predicted permeability remains approximately twice the measured value, a discrepancy not accounted for by these uncertainties. The separate model-form uncertainty associated with the conductance formulation can reach a factor of 2.6 (Appendix~\ref{app:conductance}). As this acts predominantly as a common shift in the absolute permeability, it does not resolve the difference in relative permeability between the samples. The model predicts a 7A-to-6B permeability ratio of 1.35, compared with 2.5 experimentally.

Varying $l_{Y,c}$ over the micrograph-constrained range of 100-150~nm gives permeability ranges of $3.1$-$6.1\cdot10^{-15}$~m$^2$ for 7A and $2.5$-$4.5\cdot10^{-15}$~m$^2$ for 6B. The measured value for 7A lies within its predicted range, whereas that for 6B is approximately 11\% below the lower bound. Thus, the uncertainty in the coarse correlation length substantially reduces the discrepancy for 6B, although the nominal prediction range does not quite reach the measurement.

Varying the two correlation lengths independently reveals their distinct contributions to the reconstructed transport properties. Increasing $l_{Y,c}$ from $100$ to $450$~nm changes the surface area by less than $5\%$, but increases permeability by approximately an order of magnitude. Conversely, increasing $l_{Y,f}$ from $4.25$ to $12$~nm reduces the surface area by approximately $50\%$, while changing permeability by only $7\%$. The fine and coarse fields therefore govern different aspects of the pore structure. The fine field predominantly determines the interfacial area, whereas the coarse field controls hydraulic transport. This separation explains why similar BET areas do not imply similar permeabilities and why a single-field representation fails to reproduce the experimental measurements.

\begin{figure}
\centering
\includegraphics[width=0.7\textwidth]{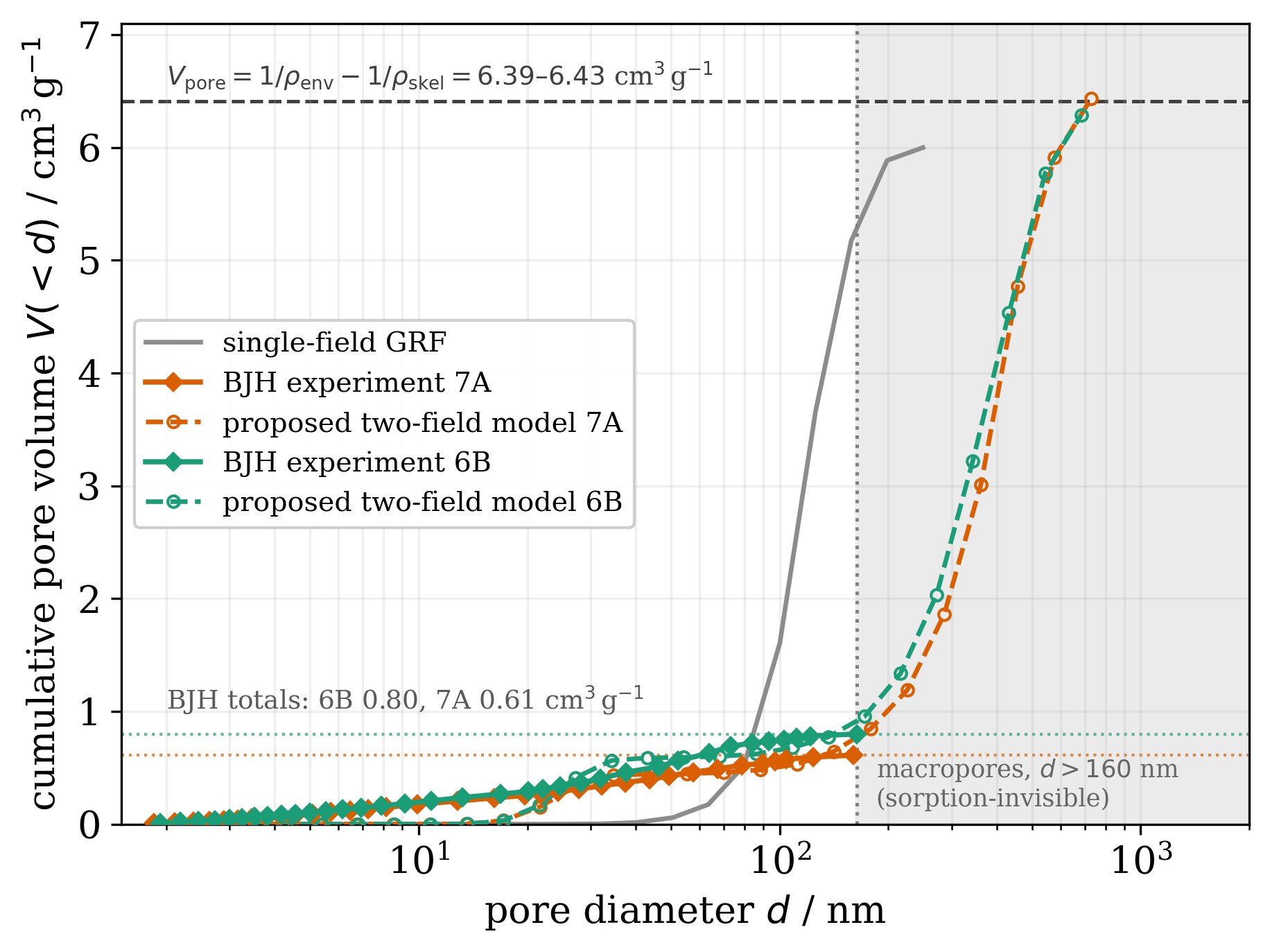}
\caption{\label{fig:psd_comparison}
Cumulative pore-size distributions of 6B and 7A from BJH measurements (filled symbols), the proposed model at $l_{Y,c} = 150$~nm (open symbols) and a single-field GRF at $\phi_0 = 0.90$, $l_Y = 40$~nm (grey). The dashed line marks the density-derived $V_{\mathrm{pore}}$ of Eq.~\eqref{eq:vtot} and the shading the pores above the sorption window.}
\end{figure}

The reconstructed surface areas are $3.5\cdot10^{7}$ and $4.1\cdot10^{7}$~m$^2$\,m$^{-3}$ for 7A and 6B, respectively, corresponding to deficits of $29\%$ and $8\%$ relative to BET measurements. These differences are consistent with the contribution of pores smaller than the voxel size (Sec.~\ref{sec:comp:convergence}). The reconstructed pore-size distributions are compared with the BJH measurements in Fig.~\ref{fig:psd_comparison}. The flux-weighted quantities discussed in Sec.~\ref{sec:results:flux} are insensitive to all three sources of uncertainty.

The pore population that dominates the internal surface area therefore need not dominate hydraulic transport. While sorption measurements constrain the fine structure, permeability is primarily sensitive to the larger-scale geometry and connectivity of the conducting pores. The latter are less directly accessible through conventional characterization, but exert the strongest influence on transport. The following section examines this distinction through the spatial distribution of flow within the reconstructed pore network.

\FloatBarrier

\subsection{\label{sec:results:flux}Flow localization and pore-network pathways}

\begin{figure*}
    \centering
    \includegraphics[width=0.95\textwidth]{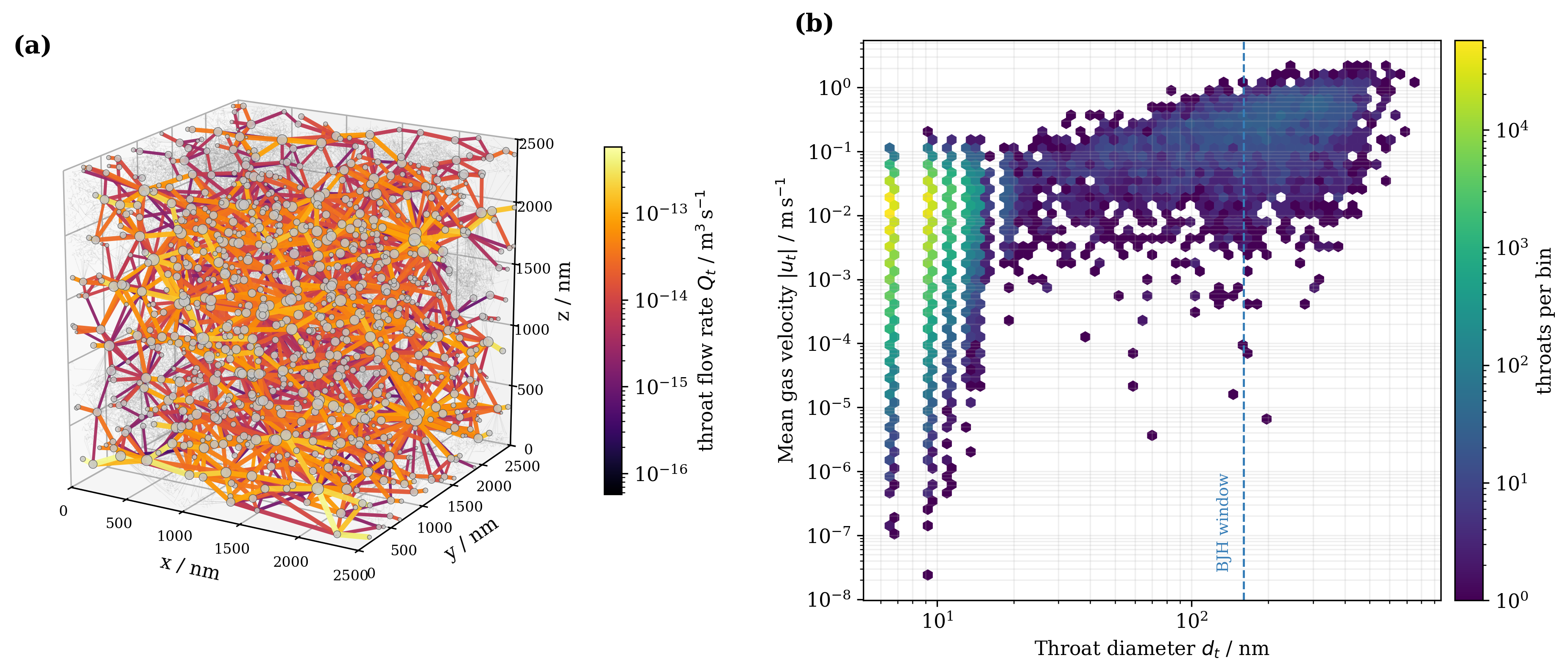}
    \caption{\label{fig:flux_weighted}
    Flow rate of the 7A network of the proposed model. (a)~Backbone of throats carrying \(90\%\) of the flux, coloured and scaled by flow rate, with its pore bodies (grey) and a random sample of the remaining throats (thin grey). (b)~Mean gas velocity per throat against throat diameter.}
\end{figure*}

We examine the distribution of gas transport across the hierarchical pore structure using the converged reconstruction of sample 7A with the proposed model at a pressure difference of \(\Delta P=50\)~mbar. The network provides access to the flow rate, mean velocity, local Knudsen number and rarefaction enhancement factor of each throat. The resulting flux distribution reveals a marked distinction between the geometrical pore population and the pathways that carry the flow (Fig.~\ref{fig:flux_weighted}). The number-weighted mean throat diameter is \(9\)~nm, compared with a flux-weighted mean of \(340\)~nm. More than \(99\%\) of the throats lie within the BJH measurement window (\(d_t<160\)~nm), but collectively carry only \(5\%\) of the total flow. The flux-weighted diameter is insensitive to grid refinement, unlike the number-weighted statistics, and thus provides a more representative measure of the transport-active pore structure (Sec.~\ref{sec:comp:convergence}).

The same distinction is evident in the rarefaction regime. The number-weighted mean Knudsen number is \(\overline{\mathrm{Kn}}=17\), whereas the flux-weighted mean is \(0.46\). The corresponding number- and flux-weighted Beskok-Karniadakis enhancement factors are \(108\) and \(3.4\), respectively. Thus, the strong rarefaction associated with the numerous fine throats has a limited influence on the macroscopic flow because these throats carry little flux. The enhancement factors at the largest Knudsen numbers should not be interpreted quantitatively, since the Beskok-Karniadakis correlation is not intended to describe Knudsen diffusion in channels with \(\mathrm{Kn}\) of order \(10\).

The spatial distribution of the flux shows that approximately \(90\%\) of the total flow passes through only \(1000\) throats, corresponding to \(0.17\%\) of the extracted network (Fig.~\ref{fig:flux_weighted}a). In comparison, the single-field structures require approximately one quarter to two fifths of their throats to carry the same fraction of the flow. The conducting macropore throats sustain velocities of order \(1\)~m\,s\(^{-1}\), whereas velocities in the mesoporous network are approximately eight orders of magnitude smaller. Flow is therefore substantially more localized in the hierarchical reconstruction.

This behaviour is consistent with the critical-path interpretation of transport introduced by \citet{ambegaokar1971} and later applied to porous media by \citet{katz1986}. The throat conductances of the hierarchical network span seven to eight orders of magnitude, and the throats carrying \(90\%\) of the total flux lie within the upper \(0.6\%\) of the conductance distribution. Rarefaction preferentially enhances transport through smaller throats and narrows this distribution, but the geometric dependence of hydraulic conductance remains dominant. The conducting backbone is consequently formed by high-conductance macropore connections. The larger pores provide the principal flow pathways, while the narrower connections between them contribute disproportionately to the hydraulic resistance, consistent with the window-opening description of aerogel networks proposed by \citet{scherer1994}. This provides a network-scale interpretation of the permeability results presented in Sec.~\ref{sec:results:validation}.

The localization of flow also constrains how rarefaction modifies the macroscopic permeability. Writing the throat conductance as \(g_{ij}=g^0_{ij}C_{r,ij}\), where \(g^0_{ij}\) is the no-slip conductance defined in Eq.~\eqref{eq:throat_conductance}, the minimum-dissipation principles evaluated using the no-slip flow field give the following bounds on the network enhancement (Appendix~\ref{app:bounds}):
\begin{equation}
    \frac{1}{\langle C_r^{-1}\rangle_w}
    \;\leq\;
    \frac{k_{\mathrm{app}}}{k_\infty}
    \;\leq\;
    \langle C_r\rangle_w,
    \qquad
    w_{ij} =
    \frac{g^0_{ij}(\Delta P^0_{ij})^2}
    {\sum_{kl}g^0_{kl}(\Delta P^0_{kl})^2},
    \label{eq:bounds}
\end{equation}
where \(w_{ij}\) is the fraction of the total no-slip viscous dissipation occurring in throat \(ij\). Although the enhancement factor spans two orders of magnitude across the full network, the bounds remain close because dissipation is concentrated in a relatively narrow subset of throats. The decomposition was performed on 14 networks, the six \(\nu = 1\) structures at \(\phi_0 = 0.70\) and \(0.90\) with \(l_Y = 20\), \(40\) and \(60\)~nm and the eight reconstructions of 7A and 6B with the proposed model at \(l_{Y,c} = 100\), \(150\), \(200\) and \(300\)~nm. For all 14 networks considered, the computed enhancement lies within these bounds, with the upper bound exceeding it by at most \(3\%\).

\begin{figure}
    \centering
    \includegraphics[width=0.5\textwidth]{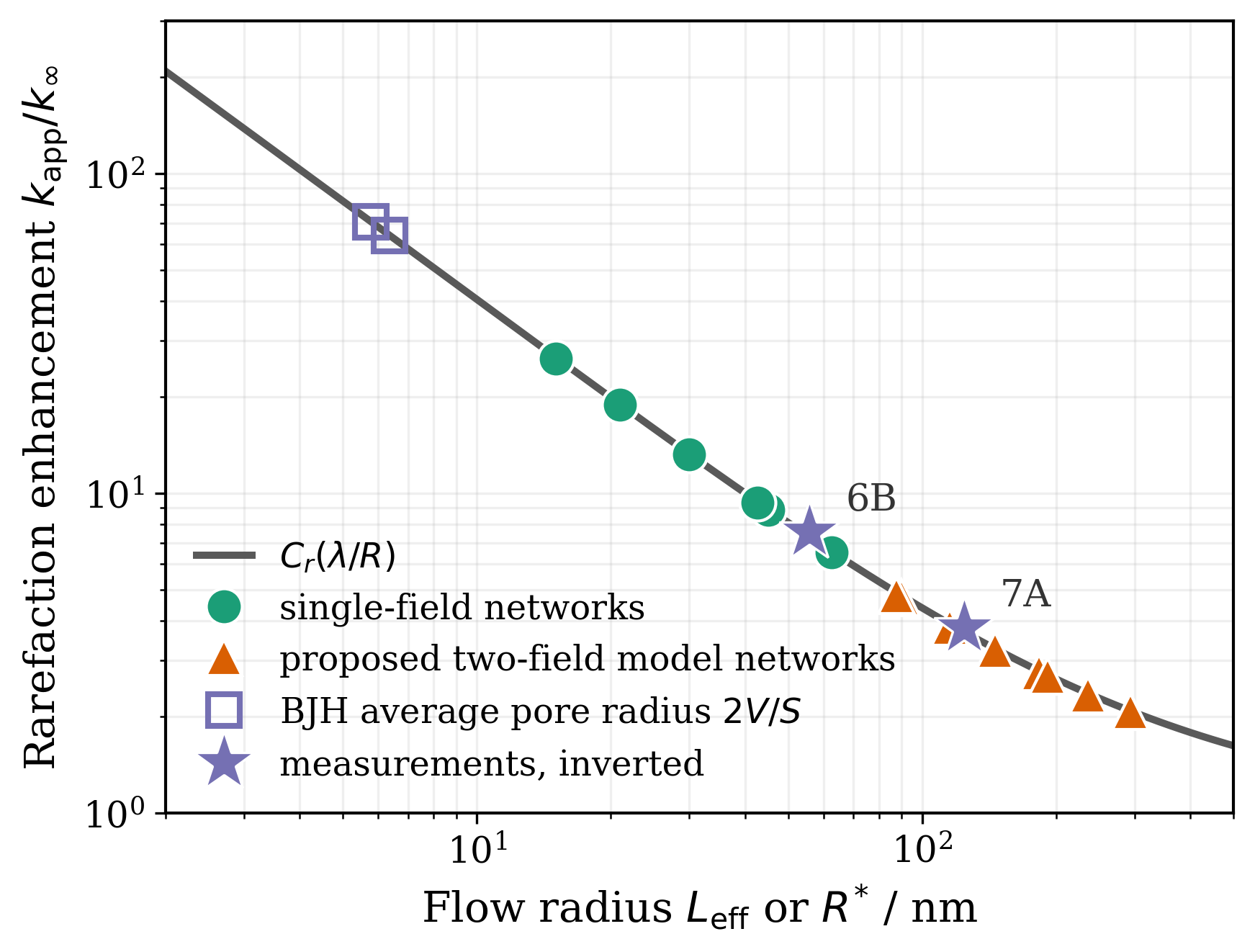}
    \caption{\label{fig:rarefaction}
    Rarefaction enhancement against the dissipation-weighted throat radius \(L_{\mathrm{eff}}\) for single-field (circles) and proposed-model (triangles) networks, with \(C_r(\lambda/R)\) (line), the conventional estimate at the BJH average pore radius \(2V/S\) of each sample (squares) and the inverted measurements at \(R^{*}\) (stars).}
\end{figure}

This result allows the network rarefaction response to be represented by a single effective throat radius,
\begin{equation}
    L_{\mathrm{eff}}=\langle R_t^{-1}\rangle_w^{-1},
\end{equation}
which is the harmonic mean of the throat radii weighted by their contribution to no-slip dissipation. Evaluating the Beskok-Karniadakis enhancement at this radius, \(C_r(\lambda/L_{\mathrm{eff}})\), reproduces the computed network enhancement to within \(3\%\) for all 14 networks (Fig.~\ref{fig:rarefaction}). Unlike a characteristic radius derived from the overall pore-size distribution, \(L_{\mathrm{eff}}\) therefore accounts for the sizes of the throats that control viscous dissipation.

The effective radius is related to the permeability-based length \(\sqrt{12k_\infty/\phi_0}\) used in porous-medium slip-flow descriptions by \citet{Lasseux2016} and \citet{Wu2017}. For the networks at \(\phi_0\approx0.9\), the relation can be written as
\begin{equation}
    L_{\mathrm{eff}} = c\sqrt{\frac{k_\infty}{\phi_0}},
\end{equation}
with \(c\) ranging from \(2.8\) to \(3.6\) and a geometric mean of \(3.1\). At \(\phi_0 = 0.7\) the three single-field networks give \(c = 3.8\) to \(4.1\) with a geometric mean of \(4.0\), so the relation is specific to the porosity.

Using \(c=3.1\), the relation between \(L_{\mathrm{eff}}\) and \(k_\infty\) allows intrinsic permeability and rarefaction enhancement to be estimated from a single gas-permeability measurement. Applied to the fourteen networks themselves, the inversion returns \(k_\infty\) within \(-22\%\) to \(+16\%\) of the no-slip value at \(\phi_0 \approx 0.9\). This is a consistency check rather than a validation, because \(c\) is calibrated on the same networks, and the band is the spread of \(c\) carried through the slip law, since changing \(c\) from \(2.8\) to \(3.6\) moves the recovered \(k_\infty\) by \(-15\%\) to \(+15\%\). The same data allow a test across families of microstructures. The value \(c = 2.9\) obtained from the six single-field networks alone, applied to the eight networks of the proposed model, recovers their \(k_\infty\) to within \(30\%\), and the value from the proposed-model networks applied to the single-field ones recovers it to within \(18\%\). At \(\phi_0 = 0.7\) the inversion with \(c = 3.1\) underestimates \(k_\infty\) by a factor of \(1.5\) to \(1.8\), and with \(c = 4.0\) recovers it to within \(8\%\). Applying the implicit relation \(k_{\mathrm{app}}=k_\infty C_r(\lambda/R^{*})\), with \(R^{*}=c\sqrt{k_\infty/\phi_0}\), to measurements near \(1\)~bar gives intrinsic permeabilities of \(1.5\cdot10^{-15}\)~m\(^{2}\) for 7A and \(3.0\cdot10^{-16}\)~m\(^{2}\) for 6B. The corresponding enhancement factors are \(3.8\) and \(7.5\), with effective flow radii of \(125\) and \(56\)~nm, respectively (Fig.~\ref{fig:rarefaction}). The inferred intrinsic permeability of 7A is close to the no-slip permeability of its reconstruction at \(l_{Y,c}=150\)~nm, \(1.9\cdot10^{-15}\)~m\(^{2}\). The inferred flow diameters of 7A and 6B are approximately \(250\) and \(110\)~nm, respectively. While the former lies above the BJH measurement window, the latter falls within its upper range, consistent with the larger BJH pore volume of 6B for pores wider than \(50\)~nm (\(0.29\) compared with \(0.19\)~cm\(^{3}\,\mathrm{g}^{-1}\) for 7A). Accounting for rarefaction increases the inferred permeability contrast between the samples from a factor of \(2.5\) to \(5\), as the narrower flow paths in 6B experience greater enhancement. These estimates depend on the network-derived constant \(c\) and should therefore be regarded as model-based inferences rather than independent determinations of intrinsic permeability. 

\FloatBarrier

\begin{table}
\begin{center}
\begin{tabular}{lcc}
\hline
Estimate & 6B & 7A \\
\hline
Mesopores only, Hagen-Poiseuille [Eq.~\eqref{eq:kmeso}]
& $2.1\cdot10^{-3}$ & $6.8\cdot10^{-4}$ \\
Mesopores only, Beskok-Karniadakis
& $0.025$ & $0.0071$ \\
Kozeny-Carman [Eq.~\eqref{eq:kc}]
& $0.035$ & $0.012$ \\
Kozeny-Carman with slip [Eq.~\eqref{eq:klinkenberg}]
& $0.27$ & $0.099$ \\
Proposed model, $l_{Y,c} = 150$~nm
& $2.0$ & $1.09$ \\
Proposed model, $l_{Y,c} = 100$--$150$~nm
& $1.12$-$2.0$ & $0.54$-$1.09$ \\
\hline
\end{tabular}
\caption{Predicted-to-measured permeability ratio 
$k_{\mathrm{model}}/k_{\mathrm{exp}}$.}
\label{tab:model_ladder}
\end{center}
\end{table}

\subsection{\label{sec:results:analytical}Estimates from sorption-derived length scales}

We first estimate the permeability associated with the pore space resolved by the BJH analysis. Treating each pore-size class as a bundle of tortuous capillaries gives
\begin{equation}
    k_{\mathrm{meso}}
    =
    \frac{1}{32\tau^2}
    \sum_i \Delta\phi_i d_i^2,
    \qquad
    \Delta\phi_i=\rho_{\mathrm{env}}\Delta V_i,
    \label{eq:kmeso}
\end{equation}
where $\Delta V_i$ is the incremental BJH pore volume in bin $i$, $d_i$ is the corresponding pore diameter, and $\tau=1.51$ is the conventional isotropic value of Eq.~\eqref{eq:tauh_affine}. The resulting permeabilities are $4.8\cdot10^{-18}$~m$^2$ for sample 6B and $3.8\cdot10^{-18}$~m$^2$ for sample 7A, approximately $470$ and $1500$ times below the measured values. The Beskok-Karniadakis correction increases these estimates by a factor of approximately eleven, but they remain lower than experiment by factors of $40$ to $140$. Surface-area-based Kozeny-Carman estimates also fall short, by factors of $28$ to $81$, or $4$ to $10$ when first-order slip is included (Table~\ref{tab:model_ladder}).

The same geometric limitation is apparent in the single-field numerical study. For all 18 parameter groups at $\nu=1$, the slip-corrected Kozeny-Carman model overpredicts the network permeability by factors of $1.4$ to $3.4$ (geometric mean $1.8$), because the hydraulic diameter inferred from specific surface area does not adequately represent the throat constrictions controlling flow.

A further estimate follows from the strut-network model developed for wet gels~\citep{scherer1994,scherer1996}. Representing the solid phase as a cubic network of cylindrical struts with radius $r$ and length $L$, the specific surface area and envelope density are approximated by
\begin{equation}
S_{\mathrm{BET}} \simeq \frac{2}{r\,\rho_{\mathrm{skel}}},
\qquad
\rho_{\mathrm{env}} \simeq \frac{3\pi r^2 \rho_{\mathrm{skel}}}{L^{2}},
\label{eq:scherer_strut}
\end{equation}
which gives $r=4.2$ and $3.8$~nm and $L=42$ and $38$~nm for samples 6B and 7A, respectively. These dimensions define an effective geometry rather than a unique reconstruction of the pore architecture, since BET surface area and density do not uniquely determine the structure. The corresponding opening radius, $r_w=(L-2r)/\sqrt{\pi}$, gives a permeability of $k=\phi_0 r_w^2/(4\kappa_w)$, with $\kappa_w\approx2.2$~\citep{scherer1994,scherer1996}. The resulting values are $3.6\cdot10^{-17}$ and $3.1\cdot10^{-17}$~m$^2$ for samples 6B and 7A, respectively, factors of $62$ and $183$ below experiment. Applying the rarefaction correction at the window scale reduces these discrepancies to factors of $2.9$ and $7.7$. The model was originally developed for wet gels and does not explicitly account for structural changes associated with drying during aerogel synthesis.

\section{\label{sec:conclusions}Conclusions}

This work investigates gas transport in highly porous RF aerogels using throat-resolved pore-network modelling with the Beskok-Karniadakis formulation to account for rarefaction effects. To represent the hierarchical pore structure, we develop a microstructural model based on the intersection of two level-cut GRFs. Comparison with classical continuum correlations shows the limitations of the Kozeny-Carman relation across the investigated porosity range. It underestimates permeability in denser structures when slip is neglected, but overestimates it at high porosities, where the hydraulic-radius assumption does not adequately represent the flow-constraining throats. Within the single-field class, changing the random-field order parameter reshapes the pore-size distribution at fixed specific surface area but changes permeability by only a factor of approximately $1.2$. Permeability within this class is therefore governed primarily by porosity and correlation length through their influence on specific surface area.

We further identify that the gas transport in RF aerogels is controlled predominantly by the macroporous network, which remains largely unresolved by nitrogen physisorption. The proposed model reproduces the measured specific surface areas within $8$ to $29\%$ and, without any parameter fitted to permeability, predict the permeability of 7A to within $10\%$ and that of 6B to within a factor of two. The agreement for 7A lies inside the $\pm25\%$ uncertainty of the method and the factor of two for 6B lies outside it. The model therefore underestimates the measured contrast between the two samples, which have similar porosity and sorption characteristics but differ in permeability by a factor of $2.5$. Predicting such sample-to-sample variations therefore requires an independent measurement of the macropore dimensions.

The reconstructed microstructures further reveal that approximately $90\%$ of the gas flow passes through only $0.17\%$ of the extracted throats. The flow-carrying throat size, its Knudsen number and the associated rarefaction enhancement are robust to grid refinement, box size and network extraction, whereas the absolute permeability carries a $\pm25\%$ discretization uncertainty and varies by up to a factor of $2.6$ between conductance formulations. Minimum-dissipation bounds show that the network rarefaction response can be represented by a single dissipation-weighted throat radius. In the hierarchical structures, this radius exceeds the BJH average pore radius by more than an order of magnitude. Consequently, rarefaction corrections based on sorption-derived pore sizes overestimate the enhancement by factors of $13$ to $34$ for the reconstructions and by factors of $8$ and $19$ for the measured samples. The same relation provides a model-based estimate of intrinsic permeability and effective flow radius from a single gas-permeability measurement, yielding flow radii of approximately $56$ and $125$~nm for samples 6B and 7A, respectively. These estimates rely on a constant derived from the network simulations and require validation through pressure-sweep measurements or independent liquid permeability measurements.

The present study nevertheless has limitations that warrant further investigation. First, the reconstructed structures are stochastic representations rather than direct reconstructions of native aerogel morphologies. Comparison with experimentally resolved three-dimensional microstructures would help assess morphological features that may not be fully captured by the GRF description, including local connectivity, pore shape and spatial heterogeneity. Second, the pore-network abstraction and semi-empirical throat-conductance models have not been benchmarked against pore-resolved simulations performed on identical geometries. The flow-carrying throats are resolved by approximately 100 voxels, suggesting that such simulations are geometrically feasible and could provide a direct assessment of the accuracy of the network representation and its conductance models. However, at a flow-weighted Knudsen number of approximately $0.5$, pore-resolved Stokes simulations with a first-order slip condition would not provide a sufficiently accurate reference. A more definitive comparison requires a kinetic description of gas transport, for example through a pore-resolved solution of the Boltzmann equation or direct simulation Monte Carlo. Such simulations would quantify the errors introduced by the pore-network abstraction and the Beskok-Karniadakis conductance correlation. Their computational cost is expected to be substantial, particularly for the three-dimensional heterogeneous structures considered here. The conductance-related uncertainty identified in this work makes this comparison particularly important.

Further experimental validation across a broader range of aerogel formulations, gases and operating pressures is also required. Finally, the rigid-skeleton assumption warrants investigation, since adsorption-induced deformation during nitrogen physisorption and pressure-induced deformation during permeametry may alter the macroporous network governing transport. Addressing these limitations would establish the range of validity of the framework and improve the predictive modelling of gas transport in porous aerogels.

\begin{bmhead}[Acknowledgements.]
The authors thank Sarah Nefzger for RF aerogel sample preparation and characterisation, Benjamin Ignatzi for the permeability experiments, Jendrik Weise for the microstructure sensitivity analysis, and Fabian Henn for the  shaping of the aerogel samples. Cedric Gommes is grateful to the Funds for Scientific Research (F.R.S.-FNRS, Belgium) for a Research Associate position. The computations were performed on the Terrabyte high-performance clusters, jointly operated by the German Aerospace Center (DLR) and the Leibniz Supercomputing Centre (LRZ).
\end{bmhead}

\begin{bmhead}[Declaration of interests.]
The authors declare that they have no competing interests.
\end{bmhead}

\begin{bmhead}[Author contributions.]
P.~Pandit: Conceptualization, Methodology, Software, Validation, Formal analysis, Visualization, Writing - original draft. R.~Kunte: Methodology, Software, Investigation. M.~Schwan: Investigation, Data Curation, Visualization, Writing - original draft. C.~J.~Gommes: Methodology, Software, Supervision, Writing - review \& editing. J.~Kowalski: Methodology, Supervision, Writing - review \& editing. A.~Rege: Conceptualization, Methodology, Supervision, Writing - review \& editing.
\end{bmhead}

\appendix
\begin{appen}
\section{Throat-scale conductance formulations}
\label{app:conductance}

The pore-network model introduced in Sec.~\ref{sec:comp:pnm} assigns each pore-pore connection a single cylindrical throat segment with radius $R_t$ and length equal to the centre-to-centre distance $L_{ij}$ (hereafter referred to as the throat-only formulation). This appendix compares this approach with two conventional conduit formulations and evaluates their suitability for highly porous aerogel networks.

In the conduit formulation, each connection is represented by three cylindrical resistances in series, consisting of two pore-body contributions and one throat contribution,
\begin{equation}
    \frac{1}{g_{ij}}
    =
    \frac{1}{g_{p1}}
    +
    \frac{1}{g_t}
    +
    \frac{1}{g_{p2}},
    \qquad
    g_k =
    \frac{\pi R_k^4}{8\mu L_k}
    C_r(\mathrm{Kn}_k),
    \label{eq:app_conduit}
\end{equation}
where the pore-body lengths are approximated as $L_p=R_p$ and the throat length is defined as $L_t=L_{ij}-L_{p1}-L_{p2}$. Two variants are considered. Variant A uses the SNOW2 inscribed pore diameter, whereas variant B uses the volume-equivalent spherical pore diameter. At the high porosities considered here, both conduit formulations introduce geometric inconsistencies. For variant A, the extracted pore bodies are often comparable in size to the neighbouring throats, leading to artificial pore-body constrictions in series with the actual transport bottleneck. For variant B, the larger equivalent pore radii frequently overlap, causing the throat length to require an imposed lower bound. Consequently, the resulting conductance depends strongly on an arbitrary geometrical correction. Furthermore, applying cylindrical rarefaction corrections to large interconnected pore bodies extends the Beskok-Karniadakis model beyond the geometry for which it was derived.

\begin{table}
\begin{center}
\begin{tabular}{llcccc}
\hline
& & \multicolumn{3}{c}{$K$ (m$^2$)} & \\
\cline{3-5}
Formulation & & $l_Y=20$ & $l_Y=40$ & $l_Y=60$ & Ratio \\
\hline
A (conduit, ins.) & HP 
& $2.0\cdot10^{-17}$ & $1.2\cdot10^{-17}$ 
& $6.4\cdot10^{-18}$ & 0.32 \\
& BK 
& $6.1\cdot10^{-16}$ & $5.2\cdot10^{-16}$ 
& $3.4\cdot10^{-16}$ & 0.55 \\
B (conduit, eq.) & HP 
& $1.4\cdot10^{-16}$ & $4.6\cdot10^{-16}$ 
& $7.7\cdot10^{-16}$ & 5.6 \\
& BK 
& $2.2\cdot10^{-15}$ & $4.1\cdot10^{-15}$ 
& $5.3\cdot10^{-15}$ & 2.4 \\
C (throat-only) & HP 
& $4.3\cdot10^{-17}$ & $2.0\cdot10^{-16}$ 
& $4.5\cdot10^{-16}$ & 10.4 \\
& BK 
& $8.1\cdot10^{-16}$ & $1.9\cdot10^{-15}$ 
& $2.9\cdot10^{-15}$ & 3.6 \\
\hline
\end{tabular}
\caption{Scalar permeability of three fixed-porosity structures
($\phi_0=0.90$, $\nu=1$, $\Delta P=50$~mbar) calculated 
using the three conductance formulations with 
Hagen-Poiseuille (HP) and Beskok-Karniadakis (BK) corrections. 
The final column gives the permeability ratio between the 
largest and smallest correlation length structures.}
\label{tab:app_modes}
\end{center}
\end{table}

Figure~\ref{fig:app_parity} compares the three formulations against the no-slip Kozeny-Carman estimate for the single-field parametric ensemble. The Kozeny-Carman prediction is used only as a structural scaling reference. Variant A is nearly independent of the reference, because its pore-body constrictions offset the growth of the throats. Variant B lies above the reference by a factor of five on average, increasingly so for the least permeable structures, as a result of the imposed throat-length corrections. Formulation C follows the reference with the smallest scatter and preserves the expected relationship between permeability and characteristic pore size.

\begin{figure*}
\centering
\includegraphics[width=\textwidth]{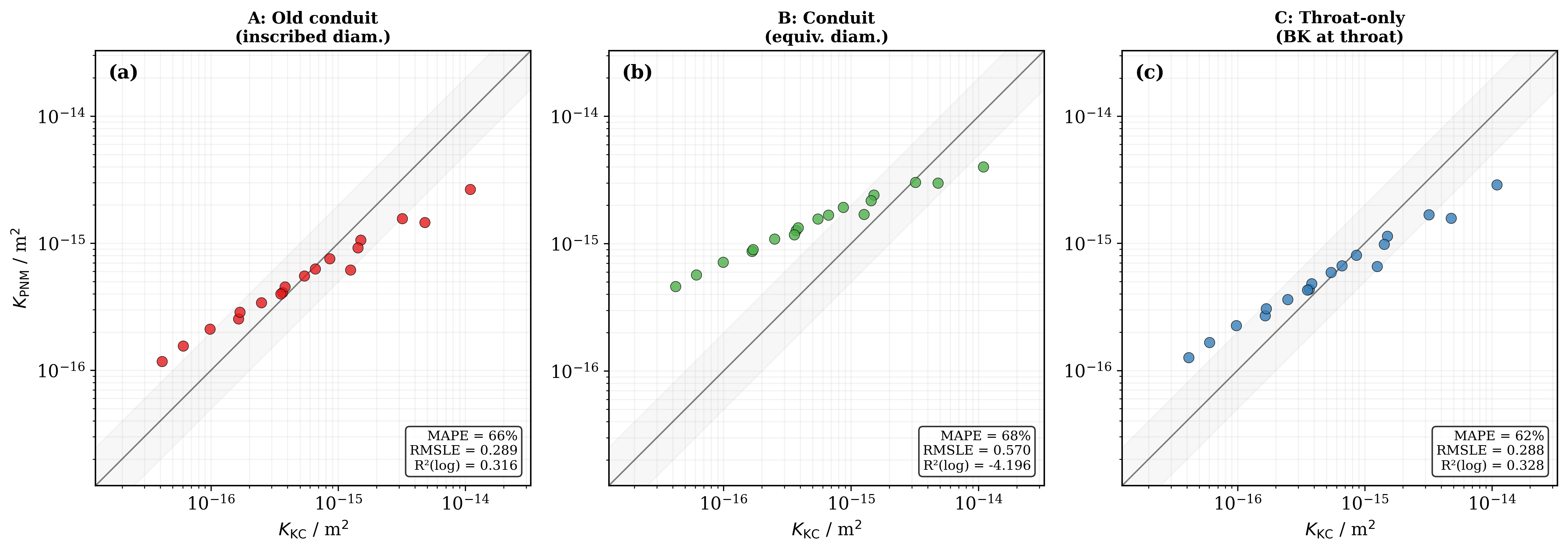}
\caption{\label{fig:app_parity}
Comparison of the three conductance formulations against the no-slip Kozeny-Carman estimate for the single-field parametric ensemble. The shaded region represents a factor of two around parity.}
\end{figure*}

The influence of the conductance formulation is further examined using three structures with identical porosity and increasing correlation length. At fixed porosity, permeability should increase with characteristic pore size. Variant A violates this expectation because the artificial pore-body constrictions become increasingly restrictive as the structure coarsens. Variant B reproduces the correct qualitative trend but systematically overestimates permeability due to the artificial extension of throat lengths. Across the 54 parametric structures, variant B predicts permeabilities a factor of 2.6 higher than the throat-only formulation on average and introduces a porosity-dependent bias. In contrast, formulation C reproduces the expected Hagen-Poiseuille scaling and captures the reduction of size sensitivity caused by rarefaction effects. The hydraulic tortuosity remains effectively unchanged between formulations.

The throat-only formulation is therefore the most physically consistent choice for the hierarchical aerogel networks considered here. By assigning the full hydraulic resistance to the narrowest resolved section, it provides a conservative estimate of throat conductance without introducing artificial pore-body restrictions or arbitrary overlap corrections. Such corrections would be substantial. The cumulative volume of the overlapping spherical pore bodies exceeds the prescribed void volume by a factor of $3.7$ at $\phi_0 = 0.85$ and $5.0$ at $\phi_0 = 0.95$, whereas the throat inscribed diameter and the pore centre-to-centre distance remain well defined throughout. Using alternative conduit formulations changes the quantitative permeability values but does not alter the central conclusions of this work. The single-field structures remain insufficient to reproduce experimental permeabilities, while only the inclusion of an additional macroporous length scale enables agreement with measurements. The spread between formulations nevertheless constitutes a model-form uncertainty of the same order as the residual model-measurement differences of Sec.~\ref{sec:results:validation}. The argument above rests on the expected trend with pore size rather than on an independent reference. A direct Stokes benchmark on the same structures, which would fix the conductance formulation, is left for future work.

\section{Microstructure model details}
\label{app:grf}
\paragraph{Level-cut Gaussian random fields.} A GRF \( Y(\mathbf{x}) \) with zero mean and unit variance is constructed as a superposition of sinusoidal components:
\begin{equation}
Y(\mathbf{x}) = \sqrt{\frac{2}{N}} \sum_{n=1}^{N} \sin\left( \mathbf{q}_n \cdot \mathbf{x} + \theta_n \right),\end{equation}
where \( \mathbf{q}_n \) are wave vectors sampled from an isotropic power spectral density, and \( \theta_n \in [0, 2\pi) \) are uniformly distributed random phases. This construction ensures statistical homogeneity and isotropy of the resulting field. A binary solid-pore structure is then obtained by defining the solid to be the points of space where the field exceeds a given threshold $a$. Formally, this corresponds to defining the indicator function of the solid as
\begin{equation}
I_S(\mathbf{x}) = H\left[ Y(\mathbf{x}) - a \right],
\end{equation}
where \( H[x] \) is Heaviside's step function, equal to 1 if $x > 0$ and to 0 otherwise. The threshold $a$ controls the solid volume fraction $\phi_1 = 1 - \phi_0$, where the subscripts $1$ and $0$ denote the solid and pore phases, so that $\phi_0$ is the porosity. Because $Y$ is standard normal at every point, the porosity equals $\Pr[Y < a]$, the cumulative distribution of the normal law evaluated at the threshold, so that
\begin{equation}
\phi_0 = \Pr[Y < a] = \frac{1}{2}\left(1 + \operatorname{erf}\left(\frac{a}{\sqrt{2}}\right)\right),
\label{eq:phi_a}
\end{equation}
where \( \operatorname{erf}(\cdot) \) is the error function. Equation~\eqref{eq:phi_a} fixes the threshold $a$ uniquely from the density-derived porosity.

The limit $\nu \to 0$, used for the fixed-porosity single-field set and for both fields of the proposed model, must be taken at fixed $l_Y$. At fixed $\xi$ the hyperbolic-secant factor of Eq.~\eqref{eq:gY} vanishes at every finite separation and the field degenerates to white noise. At fixed $l_Y$, instead, Eq.~\eqref{eq:ly} gives $\nu\xi \to l_Y/\sqrt{2}$, the sinc factor tends to unity, and
\begin{equation}
    {g_Y(r) \to \frac{1}{\cosh\left(\sqrt{2}\, r / l_Y\right)}.}
    \label{eq:gY_nu0}
\end{equation}
All $\nu = 0$ structures in this work are therefore specified by their correlation length $l_Y$ rather than by $\xi$.

\paragraph{Surface area of the proposed model.} The specific surface area decomposes in a similarly explicit manner. The boundary of a union $A \cup B$ consists of the part of $\partial A$ lying outside $B$ and the part of $\partial B$ lying outside $A$. By independence, a point of the fine interface lies in the coarse solid with probability $1-\phi_0^{(c)}$, and conversely, so the specific surface areas combine as 
\begin{equation}
    S = S_f\,(1 - \phi_0^{(c)}) + S_c\,(1 - \phi_0^{(f)}),
    \label{eq:unionsurface}
\end{equation}
where $S_f$ and $S_c$ denote the specific surface areas that each level-cut field would possess alone. These are known in closed form for Gaussian random fields~\citep{gommes2008structure,roberts1996composite}. The expected interface area per unit volume of the level set $\{Y_i = a_i\}$ equals the product of the mean gradient magnitude of the field and the Gaussian density of the level, which evaluates to
\begin{equation}
    S_i = \frac{2}{\pi}\, e^{-a_i^2/2}\, \sqrt{-\,g_Y''(0)}
        = \frac{2\sqrt{2}}{\pi}\, \frac{e^{-a_i^2/2}}{l_{Y,i}},
    \qquad i \in \{f, c\},
    \label{eq:S_levelcut}
\end{equation}
where the second equality uses the expansion $g_Y(r) = 1 - (r/l_Y)^2 + O(r^4)$ of Sec.~\ref{sec:comp:grf}, from which $-g_Y''(0) = 2/l_Y^2$. The surface area of a level-cut field is thus inversely proportional to its correlation length. Since $l_{Y,f} \ll l_{Y,c}$ implies $S_f \gg S_c$ at comparable thresholds, Eq.~\eqref{eq:unionsurface} is dominated by its first term, and essentially all of the surface area is contributed by the fine field. 

\paragraph{Calibrated values.} In practice both thresholds are set as empirical quantiles of the generated fields, so that the realized porosity matches the measured value exactly. For the reconstructions of Sec.~\ref{sec:results:validation}, $\phi_0^{(f)} = 0.48$ and $l_{Y,f} = 4.25$~nm for 7A and $\phi_0^{(f)} = 0.543$ and $l_{Y,f} = 4.05$~nm for 6B, which at the measured porosities of Table~\ref{tab:sample_props} give $\phi_0^{(c)} = 0.82$ and $0.79$. These values reproduce the BJH-window pore volumes and surface areas, $3.0\cdot10^{7}$ and $3.6\cdot10^{7}$~m$^2$m$^{-3}$, the latter to within $3\%$ when the surface area of the reconstruction is evaluated stereologically from the chord-length density, $S = 2P_L$, which avoids the overestimate of about $12\%$ of the marching-cubes triangulation at this resolution.

\paragraph{Solid connectivity.} The solid phase of a single level-cut field percolates only up to $\phi_0 \approx 0.85$. Labeling the solid voxels of the $768^3$ volumes by face connectivity, the largest solid cluster contains more than $99\%$ of the solid at $\phi_0 \le 0.75$, $93$ to $97\%$ at $\phi_0 = 0.80$ and $59$ to $70\%$ at $\phi_0 = 0.85$, and spans the domain in all three directions. At $\phi_0 \ge 0.90$ it holds less than $1.5\%$ of the solid and no cluster spans the domain, for every correlation length of the $\nu = 1$ grid. The $\nu = 0$ set at $\phi_0 = 0.906$ sits at the percolation threshold, with the largest cluster holding between $0.2$ and $56\%$ of the solid and spanning the domain in 6 of the 21 realizations. Because the pore-network model simulates the void phase, which remains connected throughout, the computed permeabilities are unaffected. The single-field structures at the experimental porosity are, however, not self-supporting solids, and they are bicontinuous only below $\phi_0 \approx 0.85$. The proposed model restores a connected skeleton at the same porosity. In every reconstruction of Sec.~\ref{sec:comp:structures} with $l_{Y,f} = 4$ to $5$~nm, over $l_{Y,c} = 150$ to $450$~nm (the $l_{Y,c} = 100$~nm reconstructions were not analysed for solid connectivity), the largest solid cluster holds $91$ to $97\%$ of the solid and spans the domain in all three directions. The coarsened-mesopore variant ($l_{Y,f} = 12$~nm) behaves alike in two of its three realizations ($92$ and $94\%$), while the third lies close to the percolation threshold ($42\%$, spanning in two directions). At $\phi_0 = 0.906$ the hierarchical description is therefore the only one of the two that yields a connected solid together with the measured transport, a structural argument independent of the pore-volume balance of Sec.~\ref{sec:exp}.

\subsection{\label{sec:comp:nueffect}Effect of the order parameter on morphology and pore size}

The order parameter $\nu$ of Eq.~\eqref{eq:gY} sets the width of the power spectral density of the field and thereby the degree of structural order, whereas the porosity fixes the threshold and the correlation length $l_Y$ fixes the specific surface area through Eq.~\eqref{eq:S_single}, independently of $\nu$. Its effect on the microstructure was quantified on realizations at fixed $\phi_0 = 0.90$ and $l_Y = 20$~nm ($500$~nm domain, $3.1$~nm voxels) with $\nu$ varied from 0 to 1 (Fig.~\ref{fig:nueffect}). Increasing $\nu$ concentrates the spectral power around the wavenumber $2\pi/\xi$, and the structure becomes more nearly periodic. The cross-sections in Fig.~\ref{fig:nueffect}(a) change from a mixture of large irregular solid clusters and small fragments at $\nu = 0$ to solid features of nearly uniform size at $\nu = 1$, and the autocovariance of the solid indicator in Fig.~\ref{fig:nueffect}(c) develops the damped oscillation of the underlying field, with a negative lobe near $r \approx 2\,l_Y$ and a weak second maximum near $3\,l_Y$. The size distributions narrow accordingly. The mean local-thickness pore diameter decreases from $66$ to $56$~nm, a reduction of $16\%$ at constant porosity, and its coefficient of variation from $0.31$ to $0.18$ [Fig.~\ref{fig:nueffect}(b)]. The solid chord-length distribution in Fig.~\ref{fig:nueffect}(d) keeps its mean at $20$~nm, in agreement with the stereological value $4\phi_1/S = 19$~nm, while its coefficient of variation falls from $0.68$ to $0.53$, and the specific surface area of the discretized structures is constant to within $3\%$, as Eq.~\eqref{eq:S_single} requires. The wider sweep of Fig.~\ref{fig:nueffect}(e), which covers $\nu$ from 0 to 10 at the experimental porosity, shows that the median pore diameter decreases from about $3.1\,l_Y$ to $2.8\,l_Y$ between $\nu = 0$ and $1$ and then saturates near $2.6\,l_Y$, so the interval $0 \le \nu \le 1$ used in this work spans most of the accessible range.

\begin{figure*}
\centering
\includegraphics[width=\textwidth]{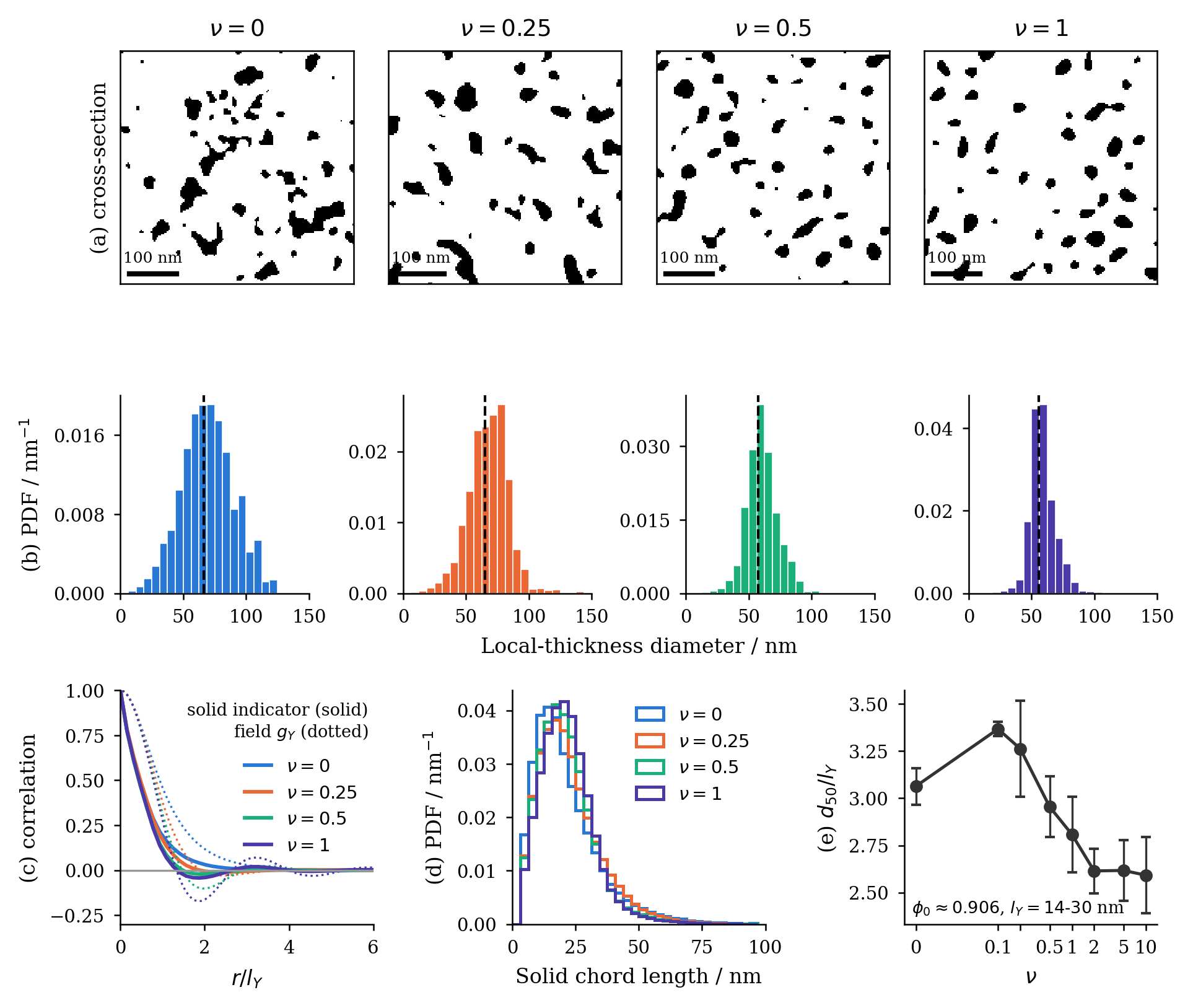}
\caption{\label{fig:nueffect} Effect of the order parameter at $\phi_0 = 0.90$ and $l_Y = 20$~nm. (a)~Mid-plane sections for $\nu = 0$, 0.25, 0.5 and 1 (solid black). (b)~Pore-diameter distributions. (c)~Solid autocovariance (solid lines) against $g_Y$ of Eq.~\eqref{eq:gY} (dotted). (d)~Solid chord-length distributions. (e)~Median pore diameter relative to $l_Y$ for $0 \le \nu \le 10$ ($\phi_0 \approx 0.906$, $l_Y = 14$-$30$~nm).}
\end{figure*}

\section{Bounds on the rarefaction enhancement of a network}
\label{app:bounds}
Consider a network with throat conductances $g_{ij}$ in which the inlet and outlet pores are held at pressures differing by $\Delta P$, and let $Q$ be the total flow. The pressure field minimizes the dissipation among all fields with these boundary values (Dirichlet principle), and the flow field minimizes it among all fluxes that conserve mass at the interior pores and carry the same total flow (Thomson principle),
\begin{equation}
    Q\,\Delta P = \min_{P} \sum_{ij} g_{ij}\,(P_i - P_j)^2,
    \qquad
    \frac{\Delta P}{Q} = \min_{q} \frac{1}{Q^{2}} \sum_{ij} \frac{q_{ij}^{2}}{g_{ij}} .
    \label{eq:app_principles}
\end{equation}
Let $g_{ij} = g^{0}_{ij}\,C_{ij}$, and let $P^{0}$, $q^{0}$ and $Q_0$ be the no-slip solution, for which $\sum_{ij} g^{0}_{ij}(\Delta P^{0}_{ij})^2 = Q_0\,\Delta P$. Inserting $P^{0}$ into the first principle gives $Q\,\Delta P \le Q_0\,\Delta P\,\langle C\rangle_w$, and inserting the rescaled flux $q^{0}Q/Q_0$ into the second gives $\Delta P/Q \le \Delta P\,\langle C^{-1}\rangle_w/Q_0$, with $w_{ij} = g^{0}_{ij}(\Delta P^{0}_{ij})^2/(Q_0\,\Delta P)$. Since $k_{\mathrm{app}}/k_\infty = Q/Q_0$ at fixed $\Delta P$, this is Eq.~\eqref{eq:bounds}. The ratio of the bounds, $\langle C\rangle_w\langle C^{-1}\rangle_w = 1 + \mathrm{Var}_w(C)/\langle C\rangle_w^{2} + \ldots$, is controlled by the spread of $C$ over the dissipating throats only. In the slip regime $C_r \simeq 1 + 4\lambda/R_t$, so that $\langle C_r\rangle_w \simeq C_r(\lambda/L_{\mathrm{eff}})$ with $L_{\mathrm{eff}} = \langle R_t^{-1}\rangle_w^{-1}$, and the Klinkenberg form $k_{\mathrm{app}} = k_\infty(1 + b_{\mathrm{K}}/p)$ follows with $b_{\mathrm{K}} = 4\lambda p/L_{\mathrm{eff}}$, a constant fixed by the dissipation-weighted throat radius. In the evaluation the mean free path is taken at the mean pressure, which over the applied pressure difference of $50$~mbar varies by less than $5\%$.

\section{Parametric transport maps}
\label{app:paramaps}

Figure~\ref{fig:app_paramaps} summarizes the transport trends obtained from the single-field parameter study discussed in Sec.~\ref{sec:results:transport}. The directional permeability components are shown in Fig.~\ref{fig:app_paramaps}(a,b). Increasing the characteristic pore size from approximately 55 to 133~nm at fixed $\phi_0=0.90$ increases permeability by a factor of about $3.5$, while the three spatial directions remain statistically equivalent with $K_{\max}/K_{\min}=1.07\pm0.04$.

Figure~\ref{fig:app_paramaps}(c) shows the pressure dependence of the apparent permeability. Increasing the applied pressure difference $\Delta P$ from 25 to 300~mbar, which raises the mean pressure in the network, reduces the apparent permeability by 10\% on average due to the decreasing molecular mean free path. The intrinsic permeabilities obtained directly from the no-slip networks ($C_r = 1$) lie between $9.5\cdot10^{-18}$ and $9.0\cdot10^{-16}$~m$^2$, $5.6$ to $27$ times below the apparent values. A Klinkenberg extrapolation over this narrow pressure range (mean pressures of $1.03$ to $1.16$~bar) would overestimate them by factors of $1.3$ to $3$.

Figures~\ref{fig:app_paramaps}(d) and (e) compare the network permeability with Kozeny-Carman estimates evaluated from the individual structural descriptors of each network. At lower porosities ($\phi_0=0.70$-$0.80$), the model underestimates permeability by up to a factor of 6.1 due to neglected rarefaction effects. At the highest porosity ($\phi_0=0.95$), it overestimates permeability by up to a factor of $2.2$ as the hydraulic-radius approximation becomes inadequate and transport approaches the behaviour of sparse fibrous networks described by drag-based models~\citep{Happel1965,duPlessis1994}. Figure~\ref{fig:app_paramaps}(f) shows the strong increase in permeability with porosity at fixed $l_Y$, with fitted exponents of $m_{\mathrm{app}}=5.3$-$5.8$ (Table~\ref{tab:scaling_summary}). These exponents reflect the coupled changes in porosity, bulk specific surface area and characteristic pore size, rather than the residual porosity dependence at fixed surface area discussed in Appendix~\ref{app:scaling}.

\begin{figure*}
\centering
\includegraphics[width=\textwidth]{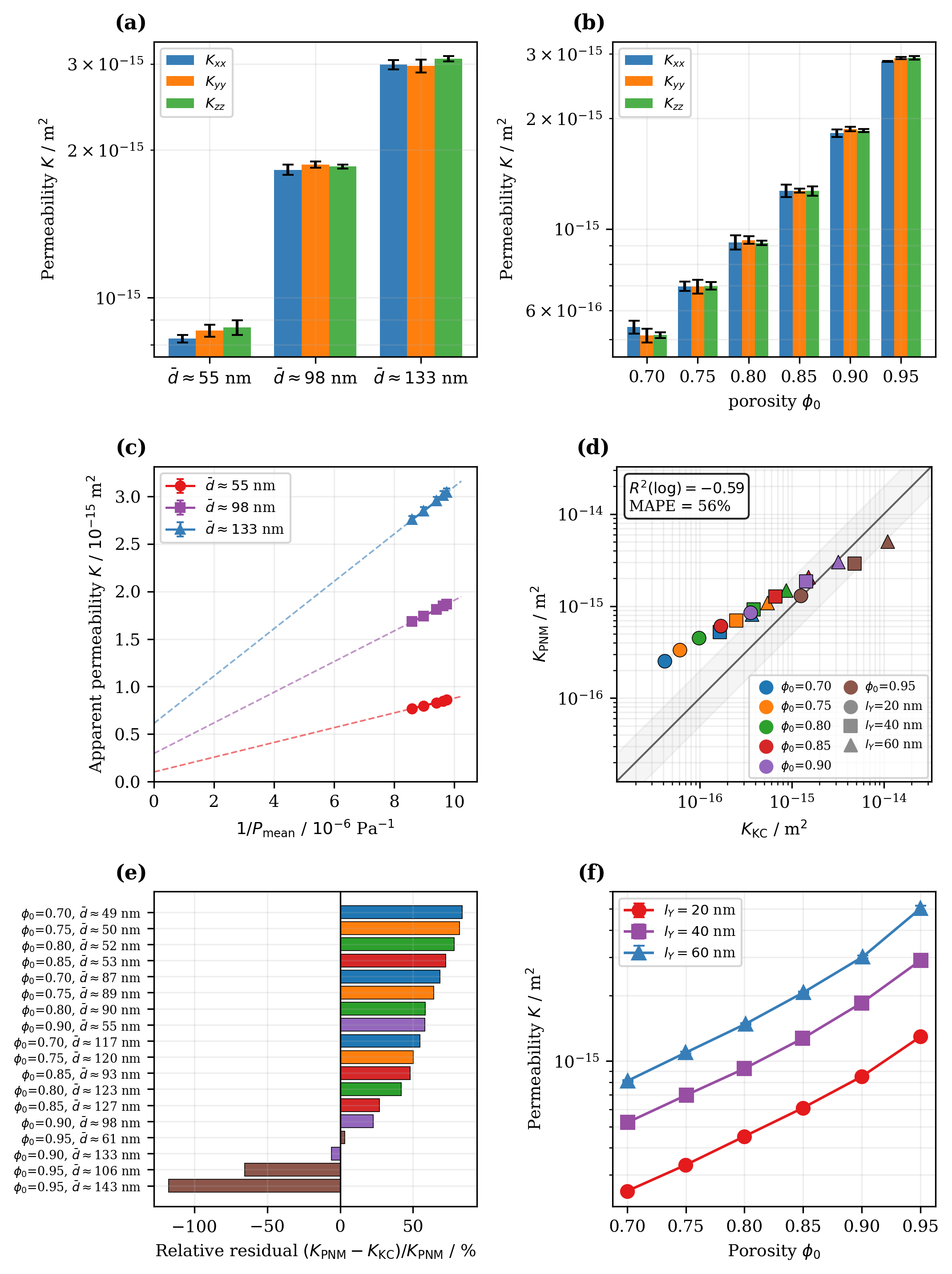}
\caption{\label{fig:app_paramaps}
Single-field transport maps. Directional permeability against (a)~mean pore diameter at $\phi_0=0.90$ and (b)~porosity at $l_Y=40$~nm. (c)~Apparent permeability against inverse mean pressure. (d)~Kozeny-Carman parity and (e)~relative residuals. (f)~Permeability against porosity.}
\end{figure*}

\section{Diffusive tortuosity computation}
\label{app:taufd}

The diffusive tortuosity $\tau_{fd}$ was calculated from a voxel-based finite-difference solution of the Laplace equation in the void phase using the PoreSpy implementation \texttt{tortuosity\_fd}. Dirichlet boundary conditions were applied on opposite faces, with the remaining boundaries treated as insulating, and the results were averaged over the three spatial directions. The linear systems were solved using algebraic multigrid. Due to sparse-matrix memory limitations, the $768^3$ domains were evaluated on centred $448^3$ subvolumes, corresponding to approximately $1460$~nm and containing several correlation lengths for all structures.

Table~\ref{tab:app_taufd} summarizes the resulting diffusive and hydraulic tortuosities. The single-field structures exhibit a narrow diffusive tortuosity range of $\tau_{fd}=1.07$-$1.09$, whereas the reconstructions with the proposed model of Sec.~\ref{sec:results:validation} lie at $1.14$-$1.19$ despite comparable hydraulic tortuosity.
\begin{table}
\begin{center}
\begin{tabular}{lcc}
\hline
Structure & $\tau_{fd}$ & $\tau_h$ \\
\hline
Single-field $l_Y{=}20$~nm & 1.086 & 1.541 \\
Single-field $l_Y{=}40$~nm & 1.077 & 1.505 \\
Single-field $l_Y{=}60$~nm & 1.074 & 1.501 \\
Single-field $l_Y{=}300$~nm & 1.071 & 1.484 \\
Proposed model 7A, $l_{Y,c}{=}150$~nm 
& $1.14 \pm 0.01$ & $1.54 \pm 0.02$ \\
Proposed model 6B, $l_{Y,c}{=}150$~nm 
& $1.19 \pm 0.05$ & $1.54 \pm 0.01$ \\
Proposed model 7A, $l_{Y,c}{=}300$~nm 
& 1.139 & 1.507 \\
Proposed model 6B, $l_{Y,c}{=}300$~nm 
& 1.148 & 1.572 \\
\hline
\end{tabular}
\caption{Diffusive and hydraulic tortuosity values. 
Diffusive tortuosity was calculated from $448^3$ subvolumes, 
while hydraulic tortuosity was obtained from the converged 
flow simulations. The proposed-model entries are the 
sorption-calibrated reconstructions of 
Sec.~\ref{sec:results:validation} at $l_{Y,c} = 150$~nm 
(mean $\pm$ standard deviation over three seeds) and the 
corresponding single realizations at $l_{Y,c} = 300$~nm.}
\label{tab:app_taufd}
\end{center}
\end{table}

\section{Single-field reference structures at the experimental porosity}
\label{app:locus}

Table~\ref{tab:app_locus} provides the seed-averaged transport properties of the single-field $\nu=0$ structures generated at the experimental porosity $\phi_0=0.906$. The 21 realizations comprise seven correlation lengths, each evaluated using three independent seeds. These structures provide the fixed-porosity reference for the permeability-surface-area relationship examined in Sec.~\ref{sec:results:envelope}. A power-law fit to the seven seed-averaged values gives
\begin{equation}
    k = 6.6\cdot10^{-5}\,S_{\mathrm{bulk}}^{-1.50},
\end{equation}
with $R^2=0.99$.

\begin{table}
\begin{center}
\begin{tabular}{lccc}
\hline
$l_Y$ (nm) & $S_{\mathrm{bulk}}$ (m$^2$\,m$^{-3}$) 
& $\bar{d}_{\mathrm{pore}}$ (nm) & $k$ (m$^2$) \\
\hline
10  & $3.8\cdot10^{7}$ & 43  & $(3.31\pm0.11)\cdot10^{-16}$ \\
15  & $2.7\cdot10^{7}$ & 54  & $(5.12\pm0.18)\cdot10^{-16}$ \\
20  & $2.0\cdot10^{7}$ & 63  & $(7.12\pm0.34)\cdot10^{-16}$ \\
50  & $8.3\cdot10^{6}$ & 111 & $(2.27\pm0.12)\cdot10^{-15}$ \\
100 & $4.1\cdot10^{6}$ & 171 & $(6.54\pm0.12)\cdot10^{-15}$ \\
200 & $2.1\cdot10^{6}$ & 237 & $(2.28\pm0.21)\cdot10^{-14}$ \\
300 & $1.5\cdot10^{6}$ & 267 & $(4.61\pm0.28)\cdot10^{-14}$ \\
\hline
\end{tabular}
\caption{Single-field GRF reference structures 
($\nu=0$, $\phi_0=0.906$; mean $\pm$ standard deviation 
over three seeds).}
\label{tab:app_locus}
\end{center}
\end{table}
\section{Scaling with surface area and porosity}
\label{app:scaling}

\subsection{Fixed-porosity scaling}
\label{app:scaling:fixedphi}

The $\nu=0$ structures at $\phi_0=0.906$ provide a direct assessment of the influence of characteristic pore size on permeability. As $l_Y$ increases from 10 to 300~nm, the local apparent surface-area exponent steepens from $-1.21$ to $-1.89$, approaching the inverse-square scaling expected in the no-slip limit. This trend reflects the decreasing influence of rarefaction as the mean Knudsen number falls from approximately 8 to 0.6. Consistently, the BK enhancement factor decreases from 30 to 1.9 over the same range.

The corresponding no-slip exponents range from $-2.2$ to $-2.6$. Their deviation from the expected value of $-2$ is partly attributable to network-extraction artefacts; when the ratio $l_Y/h$ is held fixed, the exponent approaches $-2.1$. The apparent permeability therefore approaches inverse-square scaling as the flow moves towards the continuum regime. The seed-averaged permeability data for this fixed-porosity series are given in Appendix~\ref{app:locus}, Table~\ref{tab:app_locus}. A global fit to these seven seed-averaged values gives an exponent of $-1.50$, whereas the values in Table~\ref{tab:scaling_summary} represent local slopes between neighbouring correlation lengths.

The dependence on other characteristic lengths is similarly weakened by rarefaction. At fixed porosity, the apparent permeability scales with mean pore and throat diameters with exponents of 1.34-1.57 and 1.10-1.27, respectively. The corresponding no-slip and Kozeny-Carman exponents are substantially larger. In particular, the no-slip throat-diameter exponent is close to the quadratic dependence associated with Hagen-Poiseuille flow. The exponents for all investigated scaling relations are summarized in Table~\ref{tab:scaling_summary}.

\subsection{Porosity and order-parameter effects}
\label{app:scaling:porosity}

The porosity dependence can be examined separately by writing the bulk specific surface area as
\begin{equation}
S_\mathrm{bulk}=S_0(1-\phi_0),
\end{equation}
where $S_0$ is the surface area per unit solid volume. At fixed surface area, the Kozeny-Carman relation predicts $K\propto\phi_0^3$, corresponding to a factor of approximately 2.5 between $\phi_0=0.70$ and 0.95. In contrast, the fit to the apparent permeabilities gives
\begin{equation}
K\propto S_\mathrm{bulk}^{-1.38}\phi_0^{0.61},
\end{equation}
corresponding to a variation of only approximately 1.2 over the same porosity range at fixed surface area. The corresponding no-slip fit gives a porosity exponent of $-1.0$. Thus, rarefaction substantially modifies the residual porosity dependence of the no-slip networks.

This comparison must be distinguished from varying porosity at fixed $l_Y$, for which both bulk specific surface area and characteristic pore sizes change simultaneously. The resulting apparent permeability exponents are substantially larger, as reported in Table~\ref{tab:scaling_summary}. They therefore do not represent the isolated porosity dependence at fixed surface area.

The influence of the order parameter on permeability is comparatively small. At $\phi_0\approx0.90$ and $l_Y=20$~nm, structures with $\nu=0$ and $\nu=1$ differ in permeability by approximately 20\%. Their respective mean deviations from the surface-area master curve have magnitudes of 10\% and 4\%. The associated changes in pore-size distributions and solid morphology are discussed in Sec.~\ref{sec:comp:nueffect}.

\end{appen}

\bibliographystyle{jfm}
\bibliography{apssamp}

\end{document}